\documentclass[aps,prb,twocolumn,showpacs,superscriptaddress,final,floatfix]{revtex4-1}
\usepackage[pdftex]{graphicx}  
\usepackage[update]{epstopdf}
\usepackage{amssymb}
\usepackage{amsmath}
\usepackage{color}
\usepackage{textcomp}
\usepackage[utf8]{inputenc}

\def\d{\mathrm{d}\hspace*{-0.1ex}}
\graphicspath {{./figures/}}
\makeatletter
\def\input@path{{./figures/}}
\makeatother

\begin{document}

\title{Ordering transition of the three-dimensional four-state random-field Potts model}

\author{Manoj Kumar}
 \affiliation{School of Physical Sciences, National Institute of Science Education
   and Research Bhubaneswar, Jatni 752050, India\\
   and Homi Bhabha National Institute, Training School Complex, Anushakti Nagar, Mumbai 400094, India}

\author{Martin Weigel}
\email{martin.weigel@physik.tu-chemnitz.de}
\affiliation{Institut für Physik, Technische Universität Chemnitz, 09107 Chemnitz, Germany.}

\date{\today}

\begin{abstract}
  Spin systems exposed to the influence of random magnetic fields are paradigmatic
  examples for studying the effect of quenched disorder on condensed-matter
  systems. In this context, previous studies have almost exclusively focused on
  systems with Ising or continuous symmetries, while the Potts symmetry, albeit being
  of fundamental importance also for the description of realistic physical systems,
  has received very little attention. In the present study, we use a recently
  developed quasi-exact method for determining ground states in the random-field
  Potts model to study the problem with four states. Extending the protocol applied
  for the three-state model, we use extensive finite-size scaling analyses of the
  magnetization, Binder parameter, energy cumulant, specific heat, and the connected
  as well as disconnected susceptibilities to study the magnetic ordering transition
  of the model. In contrast to the system in the absence of disorder, we find
  compelling evidence for a continuous transition, and we precisely determine the
  critical point as well as the critical exponents, which are found to differ from
  the exponents of the three-state system as well as from those of the random-field
  Ising model.
\end{abstract}

\maketitle

\section{Introduction}
\label{s1}

Random disorder is nearly unavoidable in laboratory samples of condensed-matter
systems. Understanding the effect of random doping, lattice defects and similar
impurities is hence a problem of fundamental interest for a modern theory of
condensed-matter systems \cite{dotsenko:01}. The most prominent simplified models for
such problems are systems with bond and site dilution, random-bond ferromagnets,
random-field models and spin glasses \cite{young:book}. The random-field problem, in
particular, has received much attention in the past decades, and it exhibits a rich
phenomenology with a violation of hyperscaling \cite{fisher:86a} and the curiosity of
dimensional reduction \cite{aharony:76,parisi:79a} that is present in dimensions
$d \ge 6$ (or maybe $d \ge d\mathrm{c} \approx 5$, see Ref.~\onlinecite{tissier:11}),
but broken in lower dimensions. Most effort in this direction was spent on the
random-field Ising model (RFIM), for which there is now an overall good understanding
of the phase diagram for continuous field distributions, see, e.g.,
Ref.~\onlinecite{fytas2017restoration}. For systems with continuous symmetries such
as the O($n)$ model, on the other hand, random fields destroy long-range order in
dimensions $d < 4$. \cite{imry:75,aizenmann:89a} An interesting and relevant
intermediate case are hence discrete spin systems with more than two states, such as
the Potts model \cite{wu}.

The $q$-state Potts model occupies a special place in statistical physics due to its
numerous physical realizations~\cite{wu,potts}, including soap froths, cellular
tissues, grain growth, nucleation as well as static and dynamic recrystallization,
but also for being a generalization of several special cases such as the Ising model
and random percolation. Regarding quenched disorder, the effects of bond and site
dilution as well as more general bond randomness have been relatively well studied
\cite{picco:97,ballesteros:00a,chatelain:02,berche:03a,delfino:17}. On the contrary,
randomness coupling to the order parameter, i.e., the random-field Potts model (RFPM)
that is particularly applicable to describing magnetic grains, anisotropic
orientational glasses, randomly diluted molecular
crystals~\cite{br92,michel86,*michel}, structural transitions in SrTiO$_3$
crystals~\cite{aharony}, and phase transitions in type I antiferromagnets (such as
Ndsb, NdAs, CeAs) in a uniform field~\cite{dsm}, has hardly been considered. Until
recently, there were only a few studies of this system in the
literature~\cite{shapir84,nishimori,ebjpcm, ebzpb,eb_epl,reed}, where the analytical
approaches have generally used mean-field techniques, while the computational studies
are based on Monte Carlo methods.

To illustrate the effect of the quenched random fields, consider the behavior for the
pure system. In two dimensions ($d=2$), it exhibits a continuous phase transition for
$q \le 4$ and a first-order one of $q > 4$. \cite{wu} Hence $q_\mathrm{c} = 4$ forms
a tricritical point there. For the case of three dimensions as considered in the
present paper, the transition is only continuous for $q=2$, while the three-state
model already undergoes a weak first-order transition \cite{janke:97a}. The
tricritical point has been argued to take the non-integer value
$q_\mathrm{c} \approx 2.35$ there \cite{hartmann:05}. Subjecting this system to
quenched disorder in the form of random fields, there arises the possibility of a
softening of the transitions in the first-order regime to continuous ones. As is
rigorously known \cite{aizenmann:89a}, in two dimensions this should already occur
for the smallest amount of disorder. On the contrary, in three and higher dimensions
a finite amount of disorder is in general required \cite{cardy:99a}. Also, as the
random fields yield the critical behavior of an effectively reduced dimensionality
\cite{parisi:79a,aharony:76}, one would expect the tricritical point
$q_\mathrm{c}(d)$ above which the transition becomes discontinuous to move to larger
values when random fields are included. Goldschmidt and Xu conducted a $1/q$
expansion and predict first-order transitions for $q\ge 3$, \cite{gx,*goldschmidt:86}
whereas Blankschtein {\em et al.\/} argue that both $q=3$ and $q=4$ might still
remain continuous \cite{shapir84}. Numerically, on the other hand, early results by
Reed were interpreted as indicative of a first-order transition for $q=3$, while
later Eichhorn and Binder \cite{ebjpcm,ebzpb, eb_epl} found some evidence for a
continuous transition for this case (however with a different distribution of random
fields as compared to our setup). Since these numerical studies were not yet able to
make use of advanced simulation techniques to accelerate relaxation, the rough
free-energy landscape observed for the RFPM restricted such numerical studies to very
small system sizes with the resulting strong scaling corrections. Overall the picture
of the phase diagram of the RFPM at this stage remained rather speculative.

Inspired by the substantial progress achieved for the RFIM through the use of
combinatorial optimization methods to find exact ground states
\cite{dauriac:85,ff5,gt5,kolmogorov:04}, we recently developed an algorithmic
approach suitable for finding quasi-exact ground states of samples of the RFPM for
useful system sizes \cite{kumar2018approximate}. For the random-field problem, the
renormalization-group fixed point is located at $T=0$, such that the relevant
critical behavior can be studied via the ground states. In contrast to the RFIM,
however, the ground-state problem for the RFPM with $q>2$ is non-polynomial (NP) hard
\cite{bvz,dauriac:85}. Hence an efficient exact algorithm is very unlikely to
exist. To compensate for this, we explored in Ref.~\onlinecite{kumar2023quasi} how
repeated runs of the method for $n$ different initial conditions can be used to
systematically extrapolate the approximate results in the quasi-exact limit for
$n \to \infty$.  In \cite{kumar2022critical}, we conducted an in-depth study of the
three-state RFPM in three dimensions using the quasi-exact estimates to determine its
critical behavior, finding a very clearly continuous transition with critical
exponents that are very similar to, but likely different, from those of the
three-dimensional RFIM. In the present work, we extend this line of study to the case
of the $q=4$ RFPM.  We run the ground-state method for $n = 100$ initial conditions
and extrapolate all of the disorder-averaged physical quantities to the limit of an
infinite number of initial conditions, $n \to \infty$, where the approach becomes
exact. The corresponding critical exponents are calculated using finite-size scaling
techniques.

The remainder of this paper is organized as follows. Sec.~\ref{s2} describes the
model and the numerical details of our simulations.  In Section~\ref{s3}, we present
detailed numerical results of our simulations and investigate the critical behavior
of the four-state RFPM from the approximate ground states. Finally, in Sec.~\ref{s4},
we conclude this paper with a summary and discussion.

\section{Model and methods}
\label{s2}

\subsection{Random Field Potts Model}

Depending on how the random fields couple to the spins, the Hamiltonian for the
$q$-state RFPM can take a variety of
forms~\cite{shapir84,goldschmidt:86,ebjpcm,ebzpb, eb_epl}. We consider a model where
one employs a symmetric coupling of continuous fields to each of the possible
orientations of the Potts spins~\cite{shapir84}, as follows:
\begin{equation}
\label{hamilt}
\mathcal{H}=-J\sum_{\left<ij\right>}\delta_{s_i,s_j}-\sum_i\sum_{\alpha=0}^{q-1}h_{i}^{\alpha}\delta_{s_i,\alpha}.
\end{equation}
Here, $\delta_{x,y}$ is the Kronecker delta function, and each spin $s_i$ takes one
of $q$ orientations, \emph{viz.}\ $s_i=\{0,1,....,q-1\}$. The sum across
$\langle i,j\rangle$ is over nearest neighbors $i$ and $j$ on the chosen lattice,
which is taken to be simple cubic for the present work. Periodic boundary conditions
are applied. The variables $\{h_{i}^{\alpha}\}$ denote quenched, $q$-component random
fields at site $i$, each drawn from a symmetric normal distribution, i.e.,
\begin{equation}
  P(h_i^\alpha) = \frac{1}{\sqrt{2\pi} \Delta} \exp \left[-\frac{(h_i^\alpha)^2}{2\Delta^2} \right].
  \label{eq:distrib}
\end{equation}
The width $\Delta$ of the distribution is a measure of the disorder strength in this
system.  For the special case $q=2$, it can be easily seen that the Hamiltonian
Eq.~(\ref{hamilt}) corresponds to the RFIM at coupling $J/2$ and field strength
$\Delta/\sqrt{2}$ (plus an irrelevant constant shift) \cite{kumar2018approximate}.

An alternative model Hamiltonian for the RFPM with discrete distribution of the
disorder was used in Refs.~\onlinecite{ebjpcm,gx},
\begin{equation}
  \label{hamilt_dis}
  \mathcal{H}=-J\sum_{\left< ij\right>}\delta_{s_i,s_j}-\Delta \sum_i\delta_{s_i,h_i},
\end{equation}
where the quenched random variables $h_i$ are chosen uniformly from the set
$\{0,1,\ldots,q-1\}$. At zero temperature, one has a unique ground state for the
continuous system \eqref{hamilt} \cite{vaezi:17}, while the alternative
(\ref{hamilt_dis}) might admit (extensive) degeneracies. For the RFIM, some
differences in the behavior are observed for discrete and continuous field
distributions \cite{nattermann:97} and the same might be expected here. In order to
avoid the possible subtleties associated with the existence of degeneracies, we will
focus on the form (\ref{hamilt}) of the interactions for the purpose of the present
work.

\subsection{Quasi-exact ground-state calculations}

Our study is focused on the four-state ($q=4$) RFPM in $d=3$, using simple cubic
lattices ($L^3$). In order to approximate the ground states of the considered
random-field configurations, we resort to a recently developed quasi-exact
ground-state scheme for the RFPM \cite{kumar2018approximate,kumar2023quasi}. The
basic ingredient is a technique originally developed in computer vision for the
purpose of segmenting an image with $q$ colors \cite{bvz}. Based on the observation
that multi-label problems of this type are \emph{NP} hard and hence cannot be solved
for large instances while two-color labeling can be performed efficiently using
graph-cut (or, equivalently, maximum-flow) techniques \cite{dauriac:85}, the full
$q$-color labeling is approached by randomly picking one color, and then proposing
exchanges from all other colors to the selected one and vice versa, thus effectively
embedding a solvable two-label problem into the hard $q$-color one. This approach,
known as $\alpha$-expansion \cite{bvz}, usually yields low-lying metastable
states. An adapted technique suitable for the RFPM is combined here with a systematic
use of repeated runs from random initial states in order to enable an extrapolation
of the calculation to the exact ground state \cite{kumar2023quasi}. For the sake of
brevity, we do not discuss these techniques in detail here but, instead, refer the
reader to Refs.~\onlinecite{bvz,
  kumar2018approximate,kumar2022critical,kumar2023quasi}.

\subsection{Analysis}

We performed simulations for various systems of linear size $L$ and many disorder
realizations, the exact specifications are summarized in Table~\ref{parameter}. For
each disorder configuration, we conduct simulations for $n=100$ different random
initial conditions and pick the run resulting in the lowest energy as the
ground-state estimate. For each disorder sample, we determine two basic observables:
the order parameter $m$ and the bond-energy per spin $e_J$. The former is defined as
\cite{challa1986finite}
\begin{equation}
\label{eq:mag}
 m (L,\Delta, n)= \frac{q\rho-1}{q-1},
\end{equation}
where
\begin{equation}
\label{density}
 \rho=\frac{1}{L^3} \max_{\alpha} \left(\sum_{i}\delta_{s_i,\alpha}\right)
\end{equation}
denotes the fraction of spins attaining the most common orientation, while the latter
is given by
\begin{equation}
 e_J(L,\Delta, n)= - \frac{1}{L^3}\sum_{\langle ij \rangle} \delta_{s_i,s_j}. 
\end{equation}
In a second step, by taking a disorder average $[\cdots]_{\rm av}$, we arrive at the
order parameter $[m]_{\rm av}$, the average bond-energy per spin $[e_J]_{\rm av}$,
and several other physical quantities associated with these, namely the disconnected
susceptibility $\chi_{\rm dis}$, the Binder-cumulant $U_4$, the specific heat $C$,
and the energy-cumulant $V_{4}$, which, in turn, are defined by the following
expressions:
\begin{equation}
 \chi_{\rm dis} (L,\Delta, n)= L^3[m^2]_{\rm av},
 \label{eq:chi_dis}
\end{equation}
\begin{equation}
\label{eq:bc}
U_{4} (L,\Delta, n) = 1 - \frac{[m^{4} ]_{\rm av}}{3[m^{2}]_{\rm av}^{2}},
\end{equation}
\begin{equation}
\label{cv}
C (L,\Delta, n) = \frac{\partial [e_J]_{\rm av}}{\partial \Delta},
\end{equation}
\begin{equation}
\label{ve}
V_{4} (L,\Delta, n) = 1 - \frac{[e_J^{4} ]_{\rm av}}{3[e_J^{2}]_{\rm av}^{2}}.
\end{equation}
These quantities predict the critical behavior of the system near the transition,
see, e.g., Refs.~\onlinecite{challa1986finite,chen95}.

The error bars in each observable are calculated via the jackknife method applied
over the set of disorder samples \cite{efron82, miller74}. For determining the
critical exponents, in some cases we use scaling collapses performed using the tool
\texttt{autoscale.py}, developed by O. Melchert~\cite{melchert09}. This program uses
a minimization procedure to optimize the scaling parameters via a downhill simplex
algorithm~\cite{press96}.  We also determine the goodness-of-fit parameter $Q$ to
quantify the quality of fit. This is defined as the incomplete gamma function of
$\chi^2$ and the number $f$ of degrees-of-freedom~\cite{press96}:
\begin{equation}
 Q=\Gamma\left(\frac{\chi^2}{2}, \frac{f}{2}\right).
\end{equation}
$Q$ determines the probability that a value of 
\begin{equation}
 \chi^2=\sum_{i}^{N}\left(\frac{y_i-g(x_i)}{\sigma_i}\right)^2,
\end{equation}
with $N$ data points $(x_i,y_i\pm \sigma_i)$ from a fit of the function $g$ to the
data identical to or worse than the observed value should occur \emph{by chance} if
one assumes that the model is correct~\cite{press96}. According to common practice,
if $Q\gtrsim 0.1$ the goodness-of-fit is believable. If $Q\gtrsim 0.001$, the fit may
be acceptable if the errors are non-normal or have been moderately underestimated. If
$Q < 0.001$, then the fit is not acceptable, see, e.g., Ref.~\onlinecite{press96}.

\begin{table}
\centering
\caption{Minimum number $N_{\rm samp}$ of disorder samples used per lattice size. In
  the critical regime, approximately 50\% more samples were employed. The third row
  shows the values $H(L)$ of the non-zero uniform field for each system size $L$ that
  is used for explicitly breaking the symmetry for the calculation of the
  susceptibility as discussed in Sec.~\ref{sec:susceptibility}.}
\label{parameter}
\begin{ruledtabular}
\begin{tabular}{c  c c  c  c  c c  c  c c}
  $L$&16& 20 &24&32&40&48&64&80&96\\ \hline
$N_{\rm samp}/10^3 $& 50&40&35&30&20&10&8&5&3\\ 
$H(L) \times 10^2$ &8&5.72&4.36&2.83&2.02&1.54&1&0.72&0.54\\ 
 \end{tabular}
\end{ruledtabular}
\end{table}
 
\section{Simulation Results}
\label{s3}
\subsection{Extrapolation of the physical quantities}

\begin{figure*}
  \begin{center}
    \includegraphics[width=0.75\linewidth]{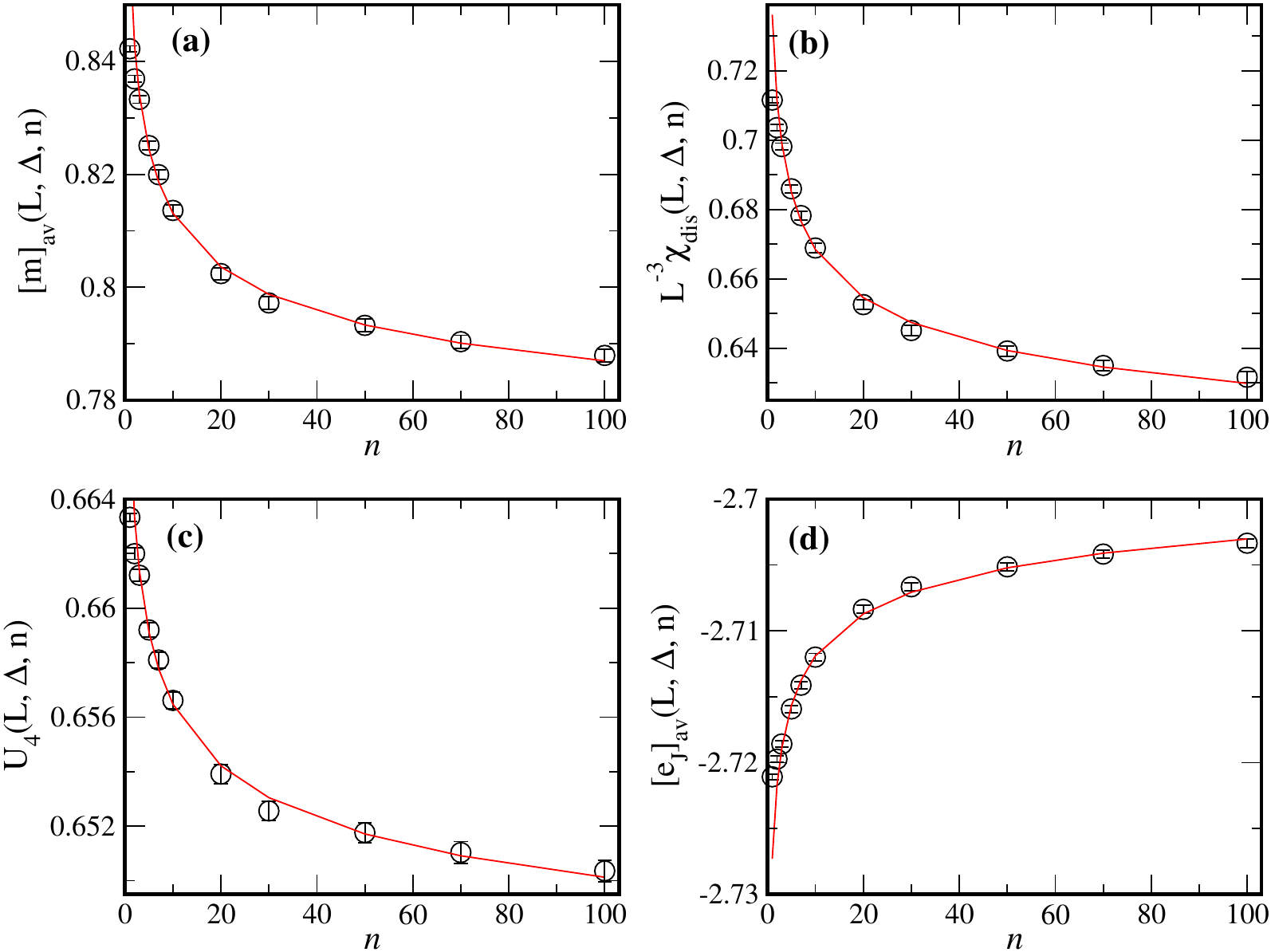}
    \caption{Disorder-averaged estimates of the magnetization $[m]_{\rm av}$, the
      disconnected susceptibility $\chi_{\rm dis}$, the Binder-cumulant $U_4$, and the
      bond-energy per spin $[e_J]_{\rm av}$ as a function of the number of initial
      conditions $n$ at a fixed value of $\Delta = 1.63$ and for lattice size $L=64$. The
      red lines show simultaneous fits to equations
      \eqref{extrapolate_m}-\eqref{extrapolate_ej} with $b=0.02$ and $e=0.242\pm0.023$. }
    \label{extrapolate}
  \end{center}
\end{figure*}

\begin{figure}
  \begin{center}
    \includegraphics[width=0.99\columnwidth]{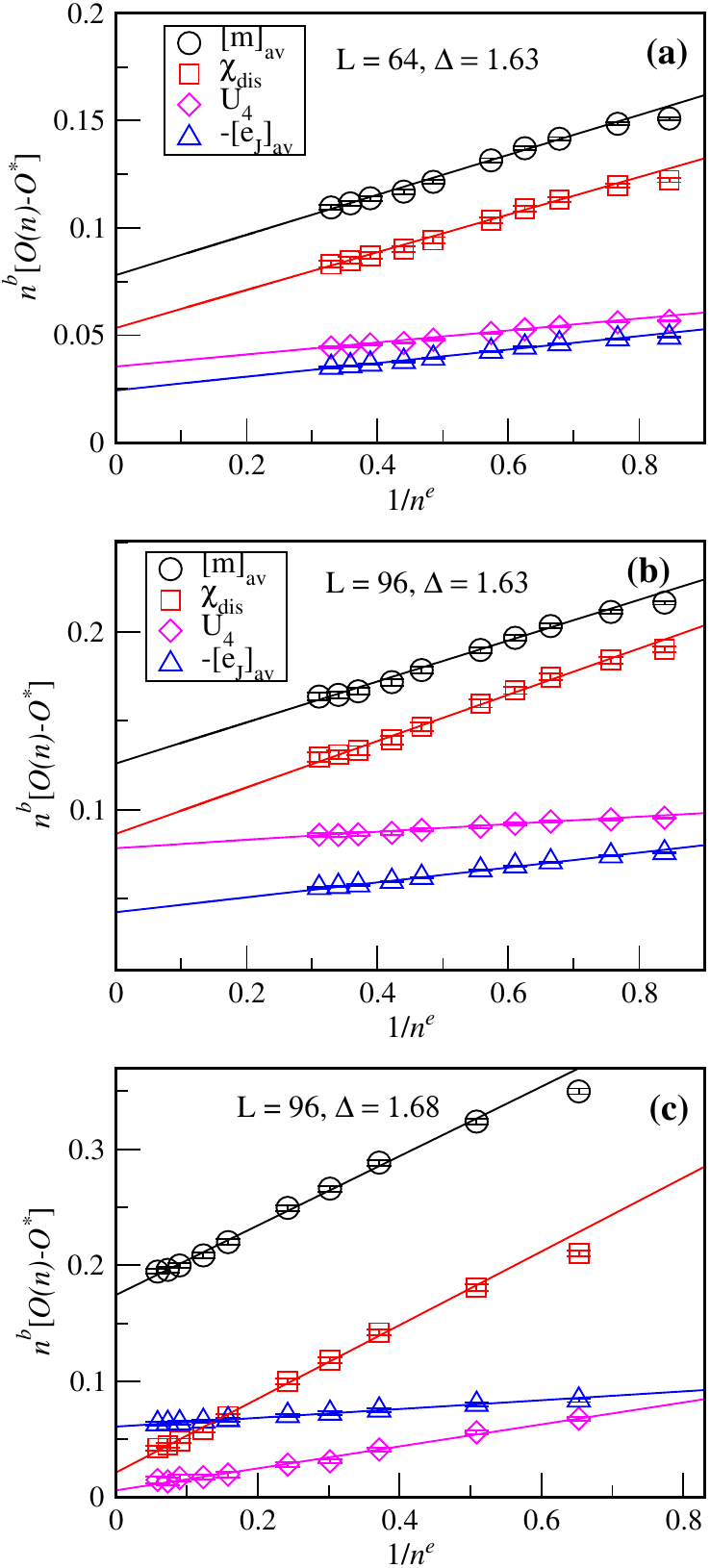}
    \caption{Residuals $n^b[O(n)-O^*]$ as a function of $1/n^e$ according to the
      scaling form \eqref{extrapolate_form} for various quantities $O$: magnetization
      $[m]_{\rm av}$, disconnected susceptibility $\chi_{\rm dis}$, Binder-cumulant
      $U_4$, and bond-energy per spin $[e_J]_{\rm av}$. Panel (a) is for $L=64$ and
      $\Delta = 1.63$, corresponding to the data in Fig.~\ref{extrapolate}. Panel (b)
      is for $L=96$ and $\Delta = 1.63$, and panel (c) is for $L=96$ and
      $\Delta = 1.68$. The solid lines are joint fits according to
      Eqs.~\eqref{extrapolate_m}--\eqref{extrapolate_ej}, which produce exponent
      estimates $e=0.242(23)$ with fit quality $Q=0.43$ for the top panel (a),
      $e=0.253(28)$ with $Q=0.81$ for the panel (b), and $e=0.60(25)$ with $Q=0.74$
      for the bottom panel (c).}
    \label{residual}
  \end{center}
\end{figure}

As outlined above, in order to improve on the results of the $\alpha$-expansion
minimization of a sample, we repeat such calculations for $n$ distinct initial
conditions of the spins and pick the run resulting in the lowest energy. To
extrapolate the results, for each disorder realization we run the simulations for
different values of $n$ up to $n_{\max}=100$ and inspect the functional form of the
dependence of numerical averages of observables on $n$.

Figure~\ref{extrapolate} shows a typical plot of different disorder-averaged
quantities (magnetization $[m]_{\rm av}$, disconnected susceptibility
$\chi_{\rm dis}$, Binder-cumulant $U_4$, and the bond-energy per spin
$[e_J]_{\rm av}$) as a function of $n$ at a fixed disorder strength $\Delta = 1.63$
and for a lattice size of $L=64$. Typically, we observe a two-stage pattern for the
convergence of these estimated zero-temperature averages. Initially, there is a fast
relaxation followed by a much slower convergence as $n$ increases. This behavior is
effectively described by the sum of two power laws~\cite{kumar2023quasi},
\begin{equation}
\label{extrapolate_form}
O(L,\Delta, n)= a n^{-b}(1+c n^{-e})+O^*(L,\Delta),
\end{equation}
where $b < e$ is the {\it asymptotic} slow exponent, $e$ represents the initial fast
decay of finite-$n$ corrections, and $O^*$ denotes the limiting value for
$n\to \infty$. As we have previously shown in Ref.~\onlinecite{kumar2023quasi}, this
form is quite generic and applies well to a specific subset of samples with known
exact ground states, which are very useful for benchmarking the approach. For these
cases we extended our study to much larger $n_{\max}=10\,000$ and found that the
residuals with respect to the exact results, i.e., $O(n)-O_{\rm ex}$, for all
considered quantities scale as $a n^{-b}(1+c n^{-e})$, with $b\simeq 0.02$ and
$e\simeq 0.5$. This behavior extends even to the cases where the exact results are
not available~\cite{kumar2023quasi, kumar2022critical}. The value of $b$ is found to
be very stable, such that we fix it to be 0.02 for the subsequent fits of the present
study, for which $n \le 100$. In order to arrive at stable results, we perform joint
fits for the different quantities with a common value of $e$, i.e.,
\begin{eqnarray}
 [m]_{\rm av}(n)&=&a_0n^{-b}(1+c_0n^{-e})+m^*, \label{extrapolate_m} \\
 \chi_{\rm dis}(n)&=&a_1n^{-b}(1+c_1n^{-e})+\chi_{\rm dis}^*, \label{extrapolate_m2}\\
 U_4(n)&=&a_2n^{-b}(1+c_2n^{-e})+U^*, \label{extrapolate_U}\\
 -[e_J]_{\rm av}(n)&=&a_3n^{-b}(1+c_3n^{-e})-e^*, \label{extrapolate_ej}
\end{eqnarray}
where $a_i$ and $c_i$, $i = 0,\ldots,3$ are amplitude parameters and $m^*$,
$\chi_{\rm dis}^*$, $U^*$, and $e^*$ are the asymptotic values of $[m]_{\rm av}$,
$\chi_{\rm dis}$, $U_4$, and $[e_J]_{\rm av}$, respectively, in the limit
$n \to \infty$.

In Fig.~\ref{residual} we show the behavior of the residuals for the different
quantities considered. According to the functional form~\eqref{extrapolate_form}, if
one plots $n^b[O(n)-O^*]$ as a function of $n^{-e}$, the data should fall on a
straight line of intercept $a$ and slope $ac$. Plotting the data in this way is hence
useful to identify any deviations from the scaling form \eqref{extrapolate_form},
especially in the asymptotic limit $n\to \infty$.  This is what we show in
Fig.~\ref{residual} for different $(L, \Delta)$ as specified, where the solid lines
are fits according to the forms \eqref{extrapolate_m}--\eqref{extrapolate_ej} which
yield fit qualities $Q = 0.43$, $0.81$, and $0.74$ for panels (a), (b), and (c),
respectively. Overall, this analysis lends credibility to the claim that the results
presented below are indeed representative of the true zero-temperature limit of the
RFPM.


In the rest of this paper, we use the extrapolated estimates together with the
estimates for finite $n$ to investigate the critical behavior via a comprehensive
finite-size scaling (FSS) analysis.

\begin{figure}
\begin{center}
\includegraphics[width=\columnwidth]{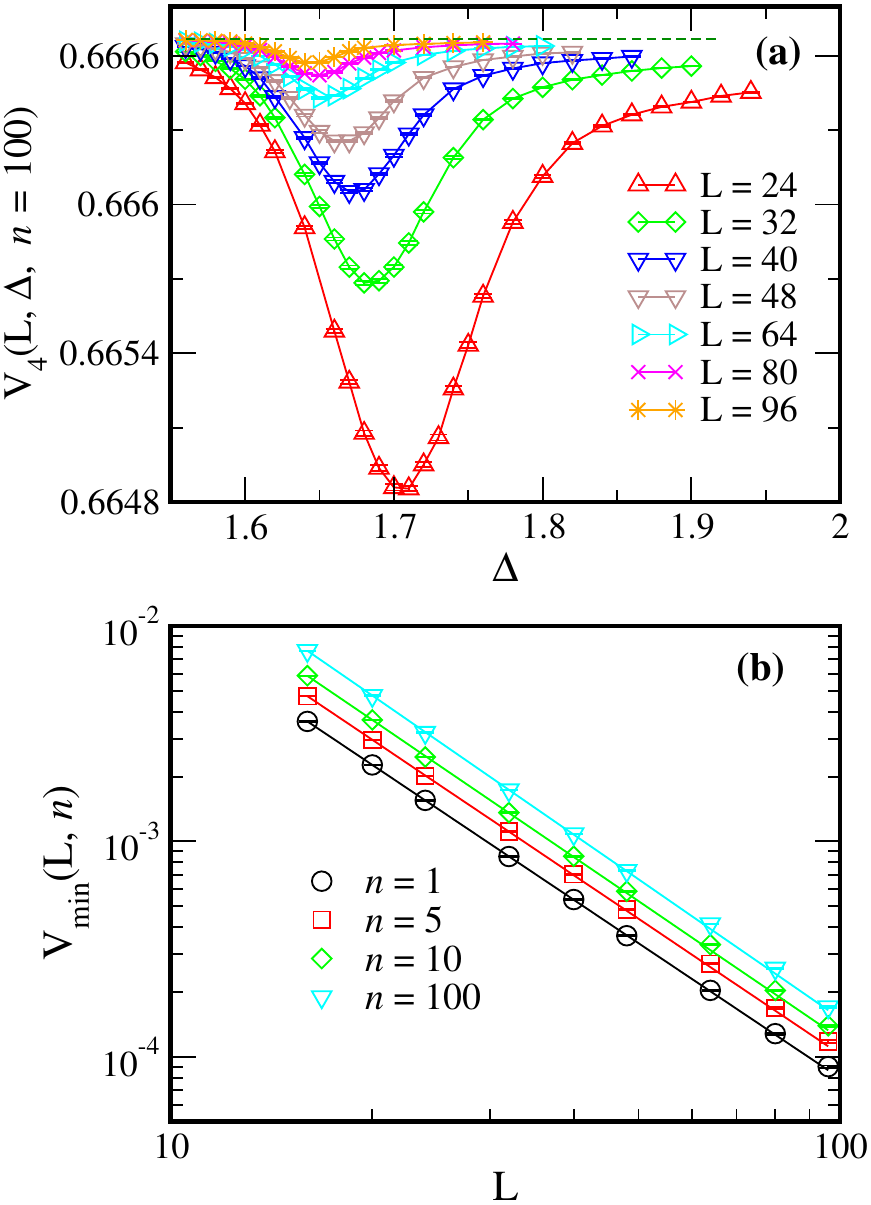}
\caption{(a) The energetic cumulant $V_4(L,\Delta)$ for $n=100$ as a function of
  $\Delta$ for various system sizes $L$. The dashed line indicates the trivial limit
  2/3 of $V_4(L,\Delta)$.  (b) $V_{\min}(L)=2/3-V_4(L,\Delta=\Delta_{\min})$ as a
  function of $L$ on a log-log scale. For a better view, the data for different $n$
  are shifted relative to each other. The solid lines are power-law fits
  $V_{\min}(L)\sim L^a$ with a value of $a \approx 2.058$. }
\label{ene_cum}
\end{center}
\end{figure}

\subsection{Energy Cumulant and Phase Transition Order}

In view of the first-order nature of the transition of the $q=4$ Potts model in the
absence of disorder and the lack of clarity from previous studies about the shift in
the tricritical point $q_\mathrm{c}$, \cite{gx,*goldschmidt:86,shapir84,ebjpcm,ebzpb,
  eb_epl} our first task is to determine the order of the transition. One useful tool
for this purpose is the fourth order energy cumulant $V_4(L, \Delta)$ as defined in
Eq.~\eqref{ve}. It's general behavior was discussed in detail in
Ref.~\onlinecite{challa1986finite}. Away from a phase transition, the probability
distribution of the energy always tends to a (single) Gaussian in the thermodynamic
limit, centered at the expected energy $\langle E\rangle$. Also, the relative width
of this Gaussian shrinks to zero, so the distribution turns into a $\delta$
function. In this case, it is easily seen \footnote{Note that the cumulant was
  \emph{not} defined using central moments.} that $V_4(L) \to 2/3$. The same is the
case for $T = T_\mathrm{c}$ at a continuous transition. In this case, the shape of
the distribution is no longer Gaussian, but its relative width still shrinks to zero
as $L\to\infty$. On the contrary, for a first order transition the limit of $V_4(L)$
is related to the distance of the ordered and disordered peaks that stays finite in
the thermodynamic limit. Hence, $V_4(L)$ approaches a non-trivial value in this
case~\cite{challa1986finite,lee1991finite}.


 

Figure~\ref{ene_cum}(a) shows the behavior of $V_4(L, \Delta)$ as a function of
$\Delta$ at $n=100$ and for different $L$. It is clearly visible that
$V_4(L, \Delta)$ displays a minimum at a certain $\Delta$, say $\Delta_{\min}$, and
the depth of the minimum strongly decreases with increasing system size $L$. To
analyze this size dependence, we performed parabolic fits near the minimum:
$V_4(\Delta)=a_0(\Delta-\Delta_{\min})^2+V_4(\Delta_{\min})$ and obtained
$V_4(\Delta_{\min})$ as the depth of the minimum of $V_4$. In Fig.~\ref{ene_cum}(b),
we plot (on a doubly-logarithmic scale) the depth of the minima after subtraction
from 2/3 as a function of $L$ and for different $n$, i.e.,
\begin{equation}
 V_{\min}(L, n) = 2/3-V_4(L,\Delta_{\min},n).
\end{equation}
Apparently, this dependence is well described by a power law,
\[
  V_{\min}(L) \sim L^{-a}
\]
as indicated by the fits drawn as solid lines. The decay exponent is found to be
$a = 2.058(8)$ for $n=1$ and $a=2.085(9)$ for $n=100$, clearly different from the
expected values $a=d=3$ for a first-order transition~\cite{challa1986finite}. From
this non-trivial value of $a$ and the convergence of $V_4$ to $2/3$ we hence conclude
that the $q=4$ RFPM undergoes a continuous transition at $T=0$.

\begin{table}[tb]
\centering
\caption{Estimates of $\Delta_\mathrm{c}$, $\nu$, and $\beta/\nu$ according to
  Eq.~\eqref{eq:mscaling} as well as $\bar{\gamma}/\nu$ according to
  Eq.~\eqref{eq:chidis-scaling} extracted from scaling collapses of the data for
  different $n$ as well as the extrapolated data for $n\to\infty$
  ($L_\mathrm{min} = 24$). $S_1$ and $S_2$ are the qualities of the collapses
  according to \eqref{eq:mscaling} and \eqref{eq:chidis-scaling}, respectively
  ($S\approx 1$ for perfect collapses).}
\label{mag_exp}
\begin{ruledtabular}
\begin{tabular}{ c  c  c  c c c c }
 
 $n$&$\Delta_\mathrm{c}$&$1/\nu$&$\beta/\nu$&$\bar{\gamma}/\nu$ &$S_1$&$S_2$\\ \hline 
 1&1.637(4)&0.734(6)&0.0556(7)&2.8902(14)&2.82&2.88 \\ 
 5&1.625(3)&0.747(5)&0.0452(8)&2.9092(13) &1.62&2.69 \\ 
 10&1.621(5)&0.740(6)&0.0438(8)&2.9123(15)&1.58&1.20 \\
 50&1.617(7)&0.726(4)&0.0422(9)&2.9153(17)&1.45&1.08 \\
100&1.615(4)&0.702(5)&0.0413(8)&2.9182(17)&1.40&0.96 \\ 
$\infty$&1.607(2)&0.645(8)&0.0393(12)&2.9215(25)&1.22&0.77 \\ 
 \end{tabular}
 \end{ruledtabular}
\end{table}

\begin{figure}
\begin{center}
\includegraphics[width=\columnwidth]{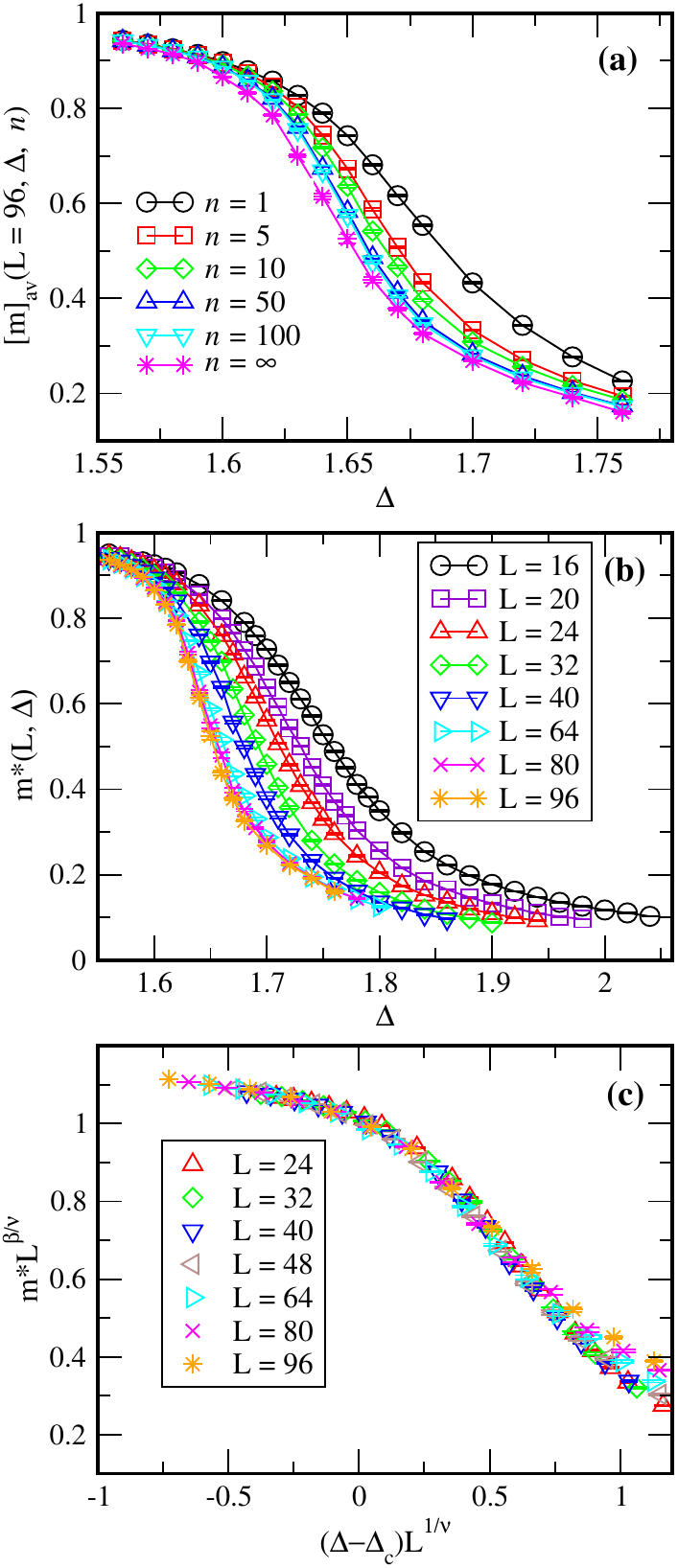}
\caption{(a) Average magnetization $[m]_{\rm av}$ as a function of $\Delta$ for
  $L=96$ and different numbers $n$ of initial conditions employed. (b) Extrapolated
  estimate $m^*$ as a function of $\Delta$ for various system sizes $L$.  (c) Scaling
  plot of $m^*(L,\Delta)L^{\beta/\nu}$ as a function of
  $(\Delta-\Delta_\mathrm{c})L^{1/\nu}$ with $ \Delta_\mathrm{c}=1.607 (2)$,
  $1/\nu=0.645(8)$, and $\beta/\nu=0.0393 (12)$.}
\label{mag}
\end{center}
\end{figure}

\subsection{The Order Parameter}

We now turn to an analysis of the critical behavior of the system at its continuous
transition. We start by considering the order parameter. Figure~\ref{mag}(a) shows
the disorder-averaged magnetization $[m]_{\rm av}$ as a function of $\Delta$ for
system size $L=96$ and a range of different numbers $n$ of initial conditions as well
as the extrapolated estimate for $n\to\infty$. As is clearly visible, $[m]_{\rm av}$
approaches the limit $n\to\infty$ rather smoothly, and we hence do not expect strong
corrections from the extrapolation procedure. In Fig.~\ref{mag}(b), we show the
lattice size dependence of the extrapolated magnetizations $m^*$ as a function of
disorder $\Delta$. The expected finite-size scaling (FSS) form of $m^*$ is
\cite{privman:privman}
\begin{equation}
 m^*(\Delta, L)= L^{-\beta/\nu}\widetilde{ \mathcal M}\left[(\Delta-\Delta_\mathrm{c})L^{1/\nu}\right].
   \label{eq:mscaling}
\end{equation}
Hence, when plotting $m^*(L,\Delta) L^{\beta/\nu}$ against
$(\Delta-\Delta_\mathrm{c})L^{1/\nu}$ with the correct values of the parameters
$\Delta_\mathrm{c}$, $\nu$ and $\beta/\nu$, the data for different $L$ should
collapse onto a single master curve near the critical region
$\Delta\approx \Delta_\mathrm{c}$, corresponding to the universal scaling function
$\widetilde{ \mathcal M}$. A correspondingly rescaled representation of the data is
shown in Fig.~\ref{mag}(c). Here, the parameters are determined using
\texttt{autoscale.py}~\cite{melchert09} with the initial choice of parameters
$\Delta_\mathrm{c}=1.62$, $1/\nu=0.7$, $\beta/\nu=0.05$ and excluding the data for
$L < L_\mathrm{min} = 24$ to effectively take scaling corrections into account. The
best collapse is obtained with the scaling parameters $\Delta_\mathrm{c}=1.607(2)$,
$1/\nu=0.645(8)$, and $\beta/\nu=0.0393(12)$ for a fitting range
$-1 \le (\Delta-\Delta_\mathrm{c})L^{1/\nu}\le 1$. We also performed a collapse
analysis for the magnetization for finite $n=1$, $5$, $10$, $50$, and $100$. In
Table~\ref{mag_exp}, we summarize the exponents from the scaling of the magnetization
for finite as well as infinite $n$ alongwith the qualities $S_1$ of the scaling
collapses \cite{melchert09}. We also attempted to extrapolate the exponent estimates
for $1/\nu$ themselves for $n\to\infty$ and found these data to be consistent with
the fits to the extrapolated magnetizations, see App.~\ref{sec:appendix}.


\begin{figure}
\begin{center}
\includegraphics[width=\columnwidth]{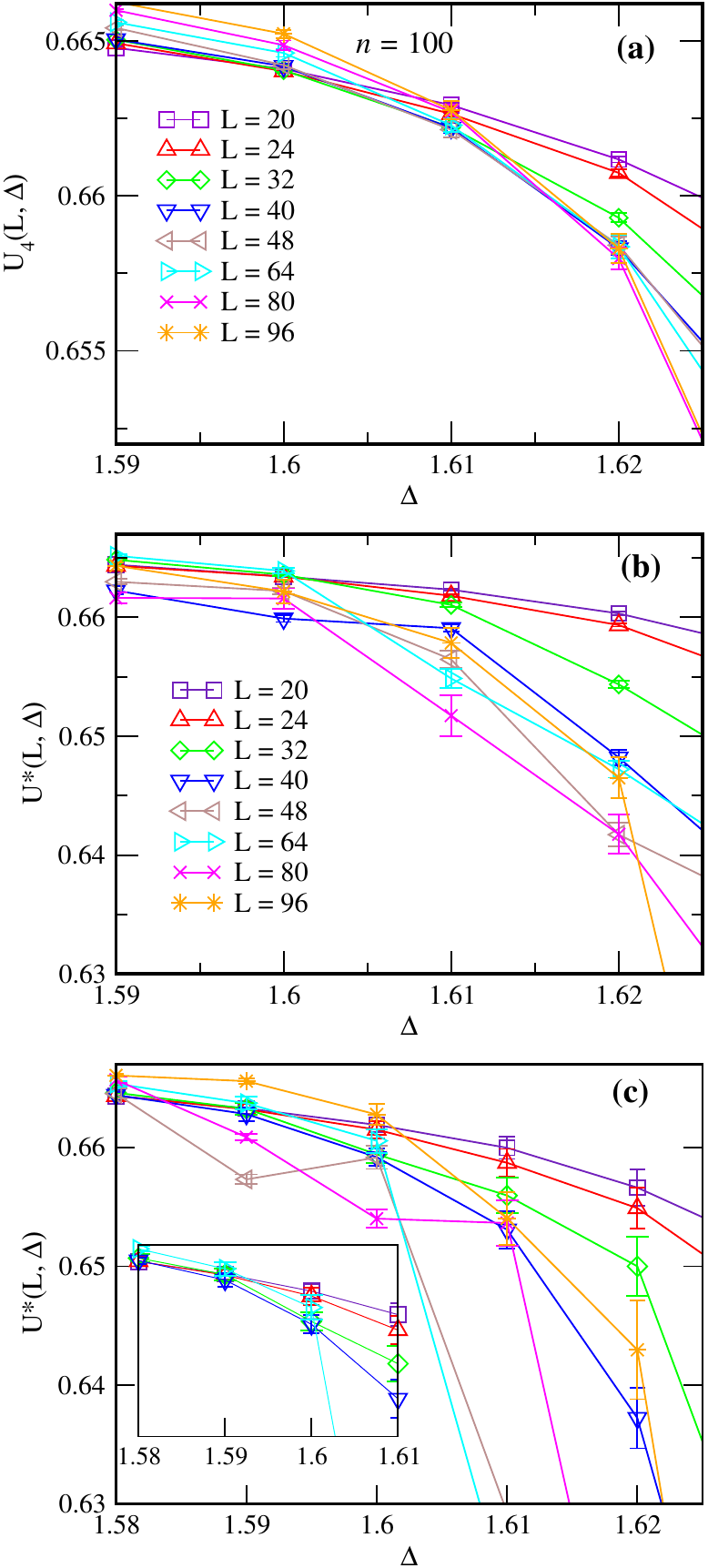}
\caption{
  (a) Binder cumulant $U_4(L,\Delta)$ for $n=100$ as a function of
  $\Delta$ for various system sizes $L$ as specified. (b) Extrapolated estimates $U^*$ of
  the Binder cumulant as a function of $\Delta$ for various system sizes $L$. (c)
  Extrapolated $U^*$ as determined from joint fits of $U_4(L,\Delta,n)$ for all $L$
  and $\Delta$ in the range 1.58 to 1.64. The joint fits are performed according to
  the extrapolated form \eqref{extrapolate_U} with the shared exponent $e$ and
  $n\ge2$.  The inset is an enlarged view of the crossing region shown for selected
  $L$.
}
\label{bc}
\end{center}
\end{figure}


\subsection{Binder Cumulant}


Next we conduct an analysis of the Binder parameter according to
Eq.~\eqref{eq:bc}. As is well known, the intersections of the cumulant curves
$U_4(\Delta, L)$ predict the location of the critical point $\Delta_\mathrm{c}$ at
which they become independent of lattice size. $L$ \cite{binder81, binder:book1} In
Fig.~\ref{bc}(a) we show the Binder cumulants $U_4(\Delta, L)$ against $\Delta$ for
$n=100$ and different linear lattice sizes $L$. The cumulant curves intersect in the
range $\Delta = (1.60,1.62)$ for $L\ge 32$, hence suggesting a corresponding location
of the critical point $\Delta_\mathrm{c}$ for $n=100$ in this area.

\begin{figure}
\begin{center}
\includegraphics[width=0.98\columnwidth]{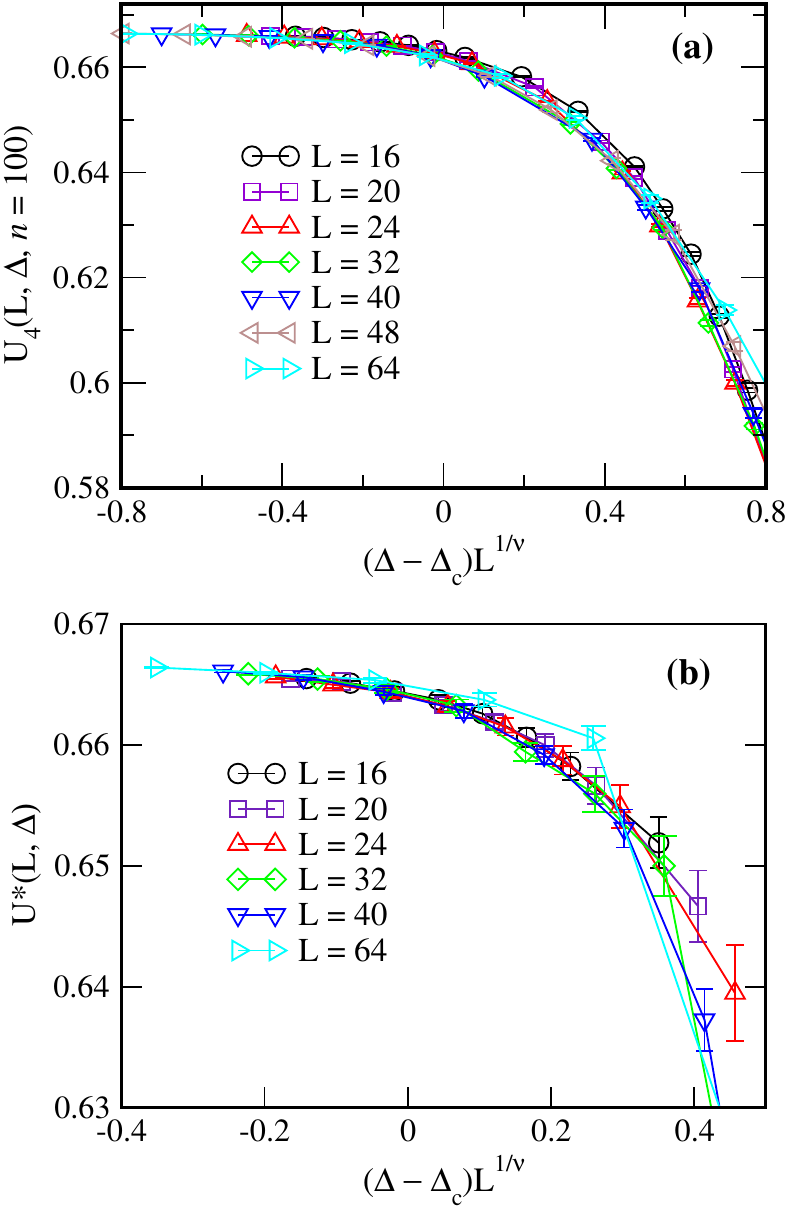}
\caption{(a) Scaling collapse of the Binder cumulant $U_4(L,\Delta)$ for $n=100$
  initial conditions as a function of $(\Delta-\Delta_\mathrm{c})L^{1/\nu}$ for system sizes in the range $L=16-64$, which yields $\Delta_c=1.6124(3)$ and $1/\nu=0.7026(16)$ with the quality of collapse parameter $S=0.74$. (b) Collapse of $U^*(L,\Delta)$ versus $(\Delta-\Delta_\mathrm{c})L^{1/\nu}$ with $\Delta_c=1.593(3)$, $1/\nu=0.655(44)$, and $S=2.34$. }
\label{bc_scale}
\end{center}
\end{figure}

In order to conduct the extrapolation for $n\to\infty$, we considered the
extrapolated results of $U_4 (L, \Delta)$ in this limit, cf.\ Fig.~\ref{bc}(b). In
this plot, however, we do not see a consistent crossing of the cumulant curves,
although a maximum crossing can be seen in the range of $\Delta = (1.6,1.61)$. To
check whether this lack of consistency arises from instabilities in the
extrapolations across different values of $L$ and $\Delta$, we attempted to
simultaneously extrapolate the cumulants for all $L$ and $\Delta$ within the range of
1.58 to 1.64 by using the extrapolated form \eqref{extrapolate_U} with a common
exponent $e$. The result of this analysis is presented in Fig.~\ref{bc}(c), showing a
clearer crossing of the cumulant curves in the range $\Delta = (1.59,1.61)$. In the
inset, an enlarged view of the crossing of $U^*(\Delta, L)$ is presented for selected
values of $L$.

The expected finite-size scaling form for $U^*(\Delta, L)=U_4(\Delta, L, n\to
\infty)$  is given by \cite{privman:privman}
\begin{equation}
 U^*(\Delta, L)= \widetilde{ \mathcal U}\left[(\Delta-\Delta_\mathrm{c})L^{1/\nu}\right],
\end{equation}
where $\widetilde{ \mathcal U}$ is a universal scaling function. We again performed a
scaling analysis using \texttt{autoscale.py} to check for consistency with this
form. The resulting collapses are displayed in Fig.~\ref{bc_scale}, where panel (a)
shows the best scaling collapse of the cumulants $U_4(\Delta, L)$ for $n=100$, which
is achieved for $\Delta_{\mathrm{c}}=1.6124(3)$ and $1/\nu=0.7026(16)$ with a quality
of collapse parameter $S=0.74$. Panel (b) shows the collapse of the extrapolated data
$U^*(\Delta, L)$ from the joint fits, yielding estimates
$\Delta_{\mathrm{c}}=1.593(3)$ and $1/\nu=0.655(44)$ with a quality of
$S=2.34$. Comparing to the data for the order parameter, we find consistent estimates
for $\Delta_c$ and $1/\nu$ for $n=100$. For the extrapolated data $1/\nu$ is also
consistent, but $\Delta_\mathrm{c}$ is slightly shifted --- an effect that we
attribute to the observed difficulties with extrapolations for this observable.

\begin{figure}[!h]
\begin{center}
\includegraphics[width=\columnwidth]{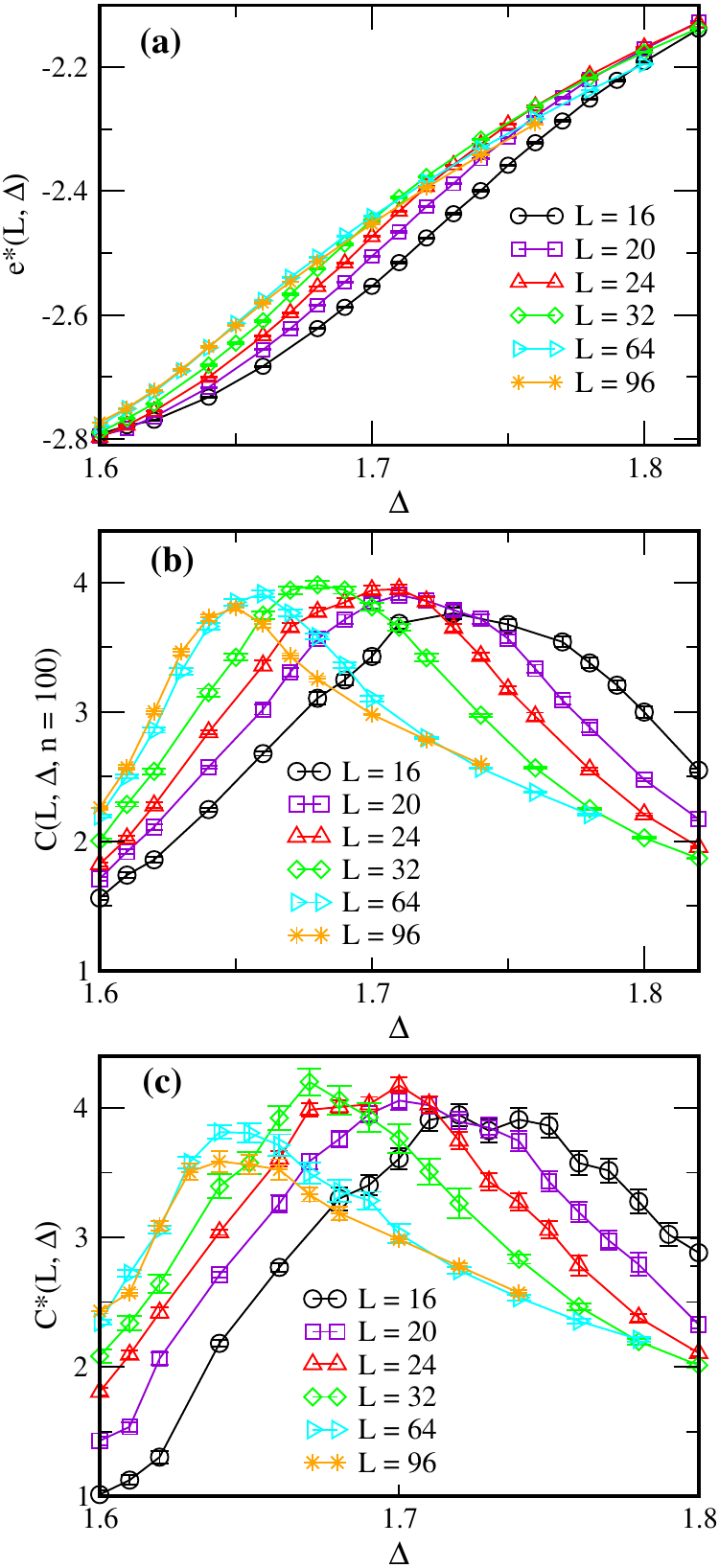}
\caption{(a) Extrapolated estimate $e^*(\Delta, L)$ of the bond energy as a function
  of $\Delta$ for various system sizes $L$.  (b) Specific heat $C(L,\Delta)$ for
  $n=100$ according to Eq.~\eqref{eq:sp_heat}.  (c) Extrapolated estimate
  $C^\ast(L,\Delta)$ of the specific heat derived by differentiating $e^*$.}
\label{sp_heat}
\end{center}
\end{figure}

\subsection{Specific Heat}

A crucial quantity in the energetic sector is the specific heat. In numerical
calculations this is usually computed from the standard fluctuation-dissipation
relation or from a temperature derivative of the internal energy. Since we operate at
zero temperature, however, these approaches are not viable here. Instead, a
specific-heat-like quantity can be obtained by differentiating the disorder-averaged
internal energy $e_J$ with respect to $\Delta$ as indicated in Eq.~\eqref{cv} (see
Ref.~\onlinecite{hartmann:01b} for details).

Numerically, we determine it using a three-point formula by taking the derivative at
an intermediate point. Since our $\Delta$ values are not equally spaced (a finer grid
is used in the vicinity of the critical point), the usual symmetric difference
formulas are not suitable. Instead, we use a three-point formula based on the Lagrange
interpolating polynomial~\cite{press96}. If $\Delta_1$, $\Delta_2$, and $\Delta_3$
are three different consecutive values of $\Delta$, an estimate of the specific-heat
$C$ at $\Delta_2$ can thus be computed as
 \begin{multline}
  \label{eq:sp_heat}
  C(\Delta_2)= \frac{(\Delta_2-\Delta_3)}{(\Delta_1-\Delta_2)(\Delta_1-\Delta_3)}[e_J(\Delta_1)]_{\rm av}\\ +\frac{(2\Delta_2-\Delta_1-\Delta_3)}{(\Delta_2-\Delta_1)(\Delta_2-\Delta_3)} [e_J(\Delta_2)]_{\rm av}\\
  +\frac{(\Delta_2-\Delta_1)}{(\Delta_3-\Delta_1)(\Delta_3-\Delta_2)}[e_J(\Delta_3)]_{\rm
    av}.
 \end{multline}

\begin{figure}
\begin{center}
\includegraphics[width=\columnwidth]{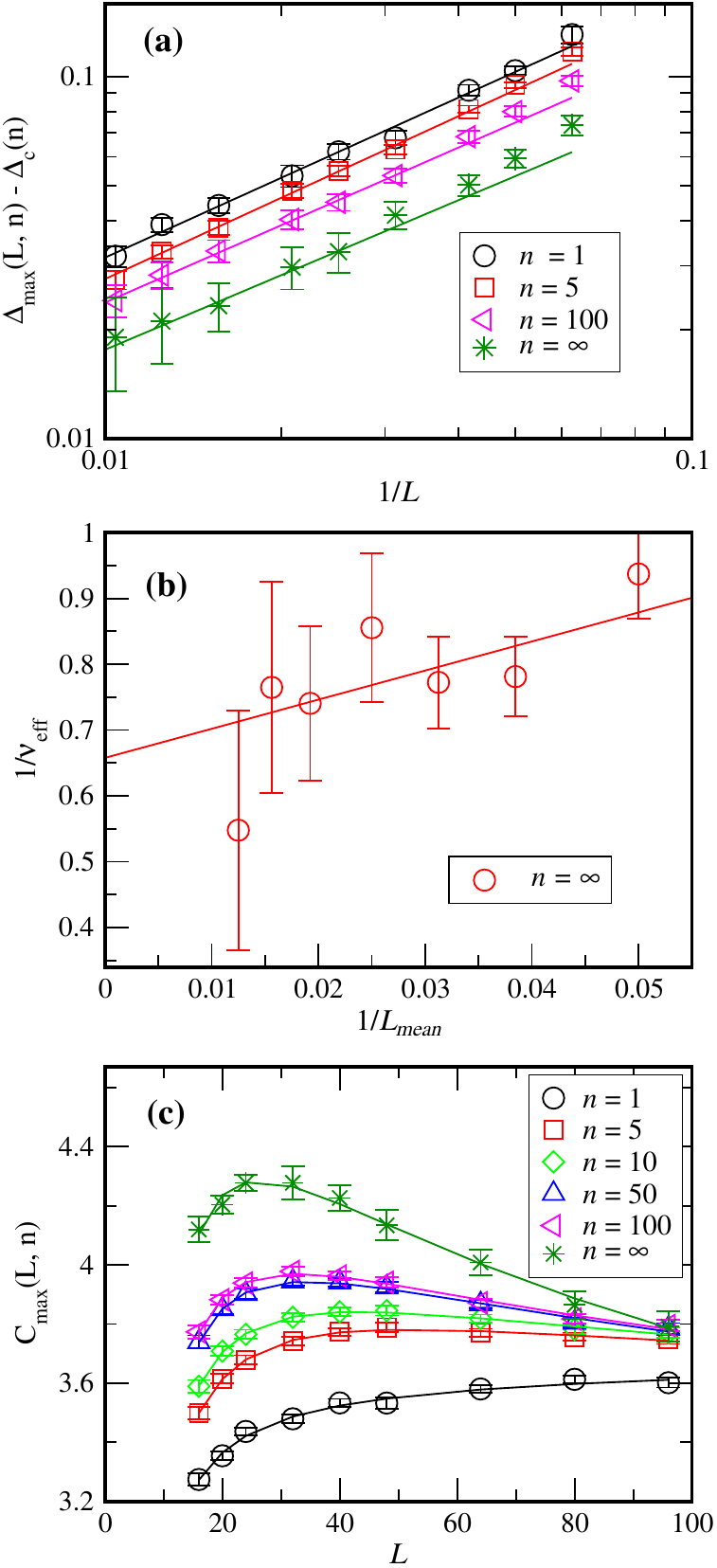}
\caption{(a) Residual peak locations $\Delta_{\rm \max}(L,n)-\Delta_c(n)$ of the
  specific heat against $1/L$ (on a double log scale) for different $n$, where
  $\Delta_c(n)$ are determined from fits of the form of Eq.~\eqref{deltac}. Solid
  lines are power-law fits, whose slope give the exponent $1/\nu$, see
  Table~\ref{sp_heat_exp}. For increased clarity, the data for different n are
  slightly shifted relative to each other. (b) Effective exponent
  $1/\nu_{\rm eff}$ as a function of $1/L_{\rm mean}$ for $n=\infty$. The solid line
  is a linear fit, yielding an extrapolated estimate $1/\nu=0.658(49)$,
  consistent with the corresponding estimate from the FSS of $m^*$, see
  Table~\ref{mag_exp}. (c) Scaling of the maxima $C_{\max}(L)$ as a function of $L$
  for different $n$. The solid curves are fits of Eq.~\eqref{cmax_scale}. The
  corresponding fit parameters $\alpha/\nu$ and $\omega$ are collected in
  Table~\ref{sp_heat_exp}. }
\label{delta_c_sp_heat}
\end{center}
\end{figure}

Using this formula, we determine the specific heat $C$ for different $L$, $\Delta$
and $n$. In panel (a) of Fig.~\ref{sp_heat}, we first present the extrapolated
estimates of the disorder-averaged bond energy per spin $e^*(L, \Delta)$ against
$\Delta$ and for different values of $L$. Figures~\ref{sp_heat}(b) and (c) display
the behavior of the specific heat as a function of $\Delta$ for different $L$, where
panel (b) shows $C(L, \Delta) $ for $n=100$, whereas panel (c) displays the
extrapolated specific heat $C^*(L, \Delta) $ after a numerical differentiation of
$e^*(L, \Delta)$ with respect to $\Delta$. In these plots, a clear peak in the
specific heat can be seen, which moves towards lower $\Delta$ with increasing $L$ and
the height of the peak initially grows with an increase in $L$. Eventually, it
decreases with $L$, indicating a negative specific heat exponent $\alpha$. To
determine the peak positions and heights, for every $L$ we performed simulations for
additional $\Delta$ values near the peak and used a parabolic fit to the peak region
of the form $C(L, \Delta)=a_0(\Delta-\Delta_{\max,C})^2+C_{\max}$ in order to obtain
the peak positions $\Delta_{\max,C}(L)$ and the peak heights $C_{\max}(L)$. The point
symbols in panels (a) and (c) in Fig.~\ref{delta_c_sp_heat}, respectively, show
$\Delta_{\max,C}(L)$ and $C_{\max}(L)$ as a function of system size $L$ for finite as
well as infinite $n$.

\begin{table}[tb!]
\centering
\caption{Parameters of fits of the forms \eqref{deltac} and \eqref{cmax_scale} to the
  locations $\Delta_{\rm \max}(L,n)$ and heights $C_{\max}(L,n)$, respectively, of
  the specific-heat peaks. $Q_1$ is the quality of the fit for
  $\Delta_{\rm \max}(L,n)$, and $Q_2$ is the quality of the fit for $C^{\max}(L,n)$.}
\begin{ruledtabular}
\begin{tabular}{ c  c  c  c c c  c}
 $n$&$\Delta_\mathrm{c}$&$1/\nu$&$\alpha/\nu$&$\omega$&$Q_1$&$Q_2$ \\ \hline
 1&1.639(10)&0.734(92)&$-0.004(56)$&$1.07(97)$ &0.03&0.42 \\ 
 5&1.633(5)&0.747(71)&$-0.058(51)$& $1.07(74)$& 0.19&0.98 \\ 
 10&1.628(4)&0.773(64)&$-0.075(52)$&$1.08(73)$ &0.14&0.97\\ 
 50&1.624(3)&0.742(54)&$-0.099(45)$&$1.18(66)$ &0.26&0.91 \\ 
100&1.621(3)&0.721(51)&$-0.088(30)$&$1.47(63)$ &0.21&0.83\\
$\infty$&1.612(6)&0.686(69)&$-0.201(75)$&$1.35(66)$&0.58&0.95\\ 
 \end{tabular}
   \end{ruledtabular} 
\label{sp_heat_exp}
\end{table}

To analyze the peak positions and heights, we note that the {\it singular} part of
the specific-heat is expected to scale as
\begin{equation}
 C_s\sim L^{\alpha/\nu}\widetilde{C}\left[(\Delta-\Delta_\mathrm{c})L^{1/\nu}\right],
\end{equation}
where $\nu$ is the correlation-length exponent and $\alpha$ is the specific-heat
exponent. At the peak, if the argument $x = (\Delta-\Delta_C)L^{1/\nu}$ of the
scaling function $\widetilde{C}$ takes some value, say $x_0$, then the peak position
$\Delta_{\max,C}(L)$ should vary as
\begin{equation}
\label{deltac}
 \Delta_{\max,C}(L)\approx\Delta_\mathrm{c}+x_0 L^{-1/\nu},
\end{equation}
and the maximum value of the singular part of the specific heat scales as 
\begin{equation}
\label{cmax_sin}
 C_{s,\max}(L)\sim L^{\alpha/\nu}.
\end{equation}
We use Eq.~\eqref{deltac} to estimate the infinite-size critical disorder
$\Delta_\mathrm{c}$ and the correlation length exponent $1/\nu$ from the data for
$\Delta_{\max,C}(L)$. The corresponding fits are shown together with the data in
Fig.~\ref{delta_c_sp_heat}(a). The resulting fit parameters for $\Delta_\mathrm{c}$
and the exponent $1/\nu$ for different values of $n$ are summarized in
Table~\ref{sp_heat_exp}, which also contains the overall satisfactory values of the
fit qualities $Q_1$. We note that the estimate of $1/\nu$ is slightly larger than the
one extracted from the order parameter, cf.\ Table~\ref{mag_exp}. In order to check
whether this is an effect of additional scaling corrections that we did not account
for, we also considered an analysis of \emph{effective exponents}. To this end, we
fixed the estimate $\Delta_\mathrm{c} = 1.612$ and performed fits of the functional
form \eqref{deltac} to the data for three consecutive lattice sizes $L_1 < L_2 < L_3$
for increasing values of $L_\mathrm{mean} = (L_1+L_3)/2$. The corresponding effective
estimates $1/\nu_\mathrm{eff}$ are shown as a function of $L_\mathrm{mean}$ in
Fig.~\ref{delta_c_sp_heat}(b). Based on general arguments of finite-size scaling,
one expects \cite{amit:book}
\begin{equation}
  1/\nu_\mathrm{eff}(L) \approx 1/\nu + b L^{-\omega},
  \label{eq:nu_eff}
\end{equation}
where $\omega$ is the Wegner correction exponent. Our data for $1/\nu_\mathrm{eff}$
are not accurate enough to allow for a determination of $\omega$, but a simple linear
extrapolation (that is consistent with the values of $\omega$ extracted from the peak
heights below, cf.\ the data in Table \ref{sp_heat_exp}) yields an estimate
$1/\nu = 0.658(49)$ that is noticeably closer to the value extracted from the scaling
of the order parameter than the uncorrected variant. This fit is also shown in
Fig.~\ref{delta_c_sp_heat}(b).  We also again considered an extrapolation of
$\nu^{-1}(n)$ for $n\to\infty$, for details see App.~\ref{sec:appendix}.

Let us now investigate $C_{\max}(L)$ as shown in Fig.~\ref{delta_c_sp_heat}(c),
which for larger values of $n$ exhibits a strong curvature (also on a logarithmic
scale), suggesting that $\alpha$ is negative. To determine $\alpha$, we initially
tried to fit the functional form
\begin{equation}
 C_{\max}(L)=C_0 + a L^{\alpha/\nu},
\end{equation}
to the data, where $C_0$ represents a non-singular background term and
$L^{\alpha/\nu}$ is the singular term as given in Eq.~\eqref{cmax_sin}.  This form
works for smaller values of $n$, but it is unable to represent the non-monotonic
behavior seen for $n\ge 10$, cf.\ Fig.~\ref{delta_c_sp_heat}(b). To describe such
effects, we need to include corrections to scaling, implementing fits of the form
\begin{equation}\label{cmax_scale}
 C_{\max}(L)=C_0 + c_1 L^{\alpha/\nu} (1+c_2L^{-\omega}),
\end{equation}
where $\omega$ corresponds to the Wegner exponent and $c_2$ is some constant. Due to
the non-linearity, we are not able to reliably fit this five-parameter form to the
data and, instead, we find that the results are heavily dependent on the initial
parameter values. For a negative $\alpha$ the maxima $C_{\max}(L)$ will approach
$C_0$ as $L \to \infty$, but a saturation to a non-zero $C_0$ is not visible for
larger $n$ as well as for $n\to\infty$ within the range of available system sizes,
cf.\ Fig.~\ref{delta_c_sp_heat}(b). In order to stabilize the fits, we might hence
assume that $C_0 = 0$, and we indeed find such fits to work quite reliably. The best
fits of this form are shown as solid curves in Fig.~\ref{delta_c_sp_heat}(b) and the
resulting estimates for the exponents $\alpha/\nu$ and $\omega$ are collected in
Table~\ref{sp_heat_exp} together with the fit qualities $Q_2$, which are found to be
excellent. Finally, concerning the possibility that $\alpha \simeq 0$, from the form
\eqref{cmax_scale} we would conclude that $C_{\max}(L) \simeq C_0' +c_2'L^{-\omega}$,
which would again imply saturation of $C_{\max}$ as $L \to \infty$. Since our data
for larger $n$ do not show any sign of saturation of $C_{\max}$, we rule out this
possibility and conclude that $\alpha$ is negative as given in
Table~\ref{sp_heat_exp}.

\begin{figure}[!h]
\begin{center}
\includegraphics[width=0.98\columnwidth]{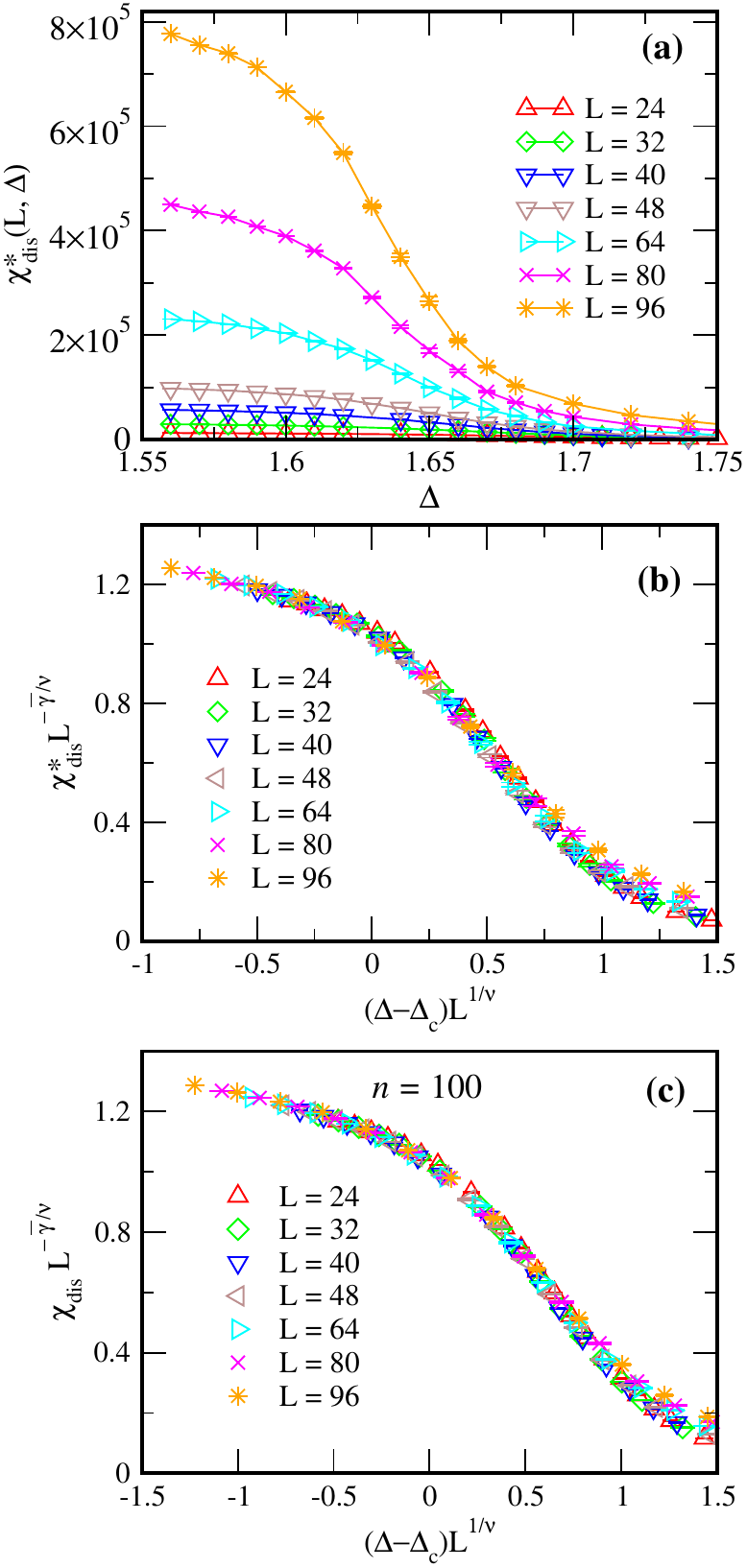}
\caption{(a) Extrapolated estimate $\chi_{\rm dis}^*(L,\Delta)$ of the disconnected
  susceptibility as a function of $\Delta$ for various system sizes $L$. Error bars
  are also shown, but they are much smaller than the symbol size.  (b) Scaling plot
  of $\chi_{\rm dis}^*(L,\Delta)L^{-\bar\gamma/\nu}$ as a function of
  $(\Delta-\Delta_\mathrm{c})L^{1/\nu}$ with the values $\Delta_\mathrm{c}$, $1/\nu$
  and $\bar \gamma/\nu$ in Table~\ref{mag_exp}. (c) Same as (b) but for $n=100$. }
\label{chi_dis}
\end{center}
\end{figure}

\subsection{Susceptibility}
\label{sec:susceptibility}

We finally considered the connected and disconnected susceptibilities.  The
\emph{disconnected} susceptibility is defined in Eq.~\eqref{eq:chi_dis}.  Panel (a)
of Fig.~\ref{chi_dis} illustrates its extrapolated estimates
$\chi_{\rm dis}^*(L,\Delta)$ as a function of $\Delta$ for different $L\geq 24$. The
expected FSS form is
\begin{equation}
  \chi_{\rm dis}^*(L,\Delta) =L^{\bar\gamma/\nu} \widetilde{\mathcal \chi}_{\rm dis}[(\Delta-\Delta_\mathrm{c})L^{1/\nu}].
  \label{eq:chidis-scaling}
\end{equation}
Based on this scaling form, we arrive at a clean scaling collapse for
$\bar\gamma/\nu=2.9215(25)$, cf.\ Fig.~\ref{chi_dis}(b). The values of
$\Delta_\mathrm{c}$ and exponent $1/\nu$ agree with those from the magnetization
scaling. We also performed FSS of the disconnected susceptibility for finite $n$ and
obtained excellent scaling collapses. As an example, in Fig.~\ref{chi_dis}(c) we show
the best scaling collapse for $n=100$.  The values of the exponent $\bar\gamma/\nu$
for different $n$, including the one for extrapolated $\chi_{\rm dis}^*(L,\Delta)$,
are listed in Table~\ref{mag_exp}, along with the qualities $S_2$ of the data
collapse. We also arrive at exponent estimates for $1/\nu$, but we find them to be
nearly indistinguishable from the values found from the magnetization (cf.\ Table
\ref{mag_exp}), so we do not list separate values here.

The \emph{connected} susceptibility $\chi (L, \Delta)$ is the response of the system
to the presence of a small uniform external field $h$. Since this study is conducted
at temperature $T=0$, we cannot use the usual fluctuation-dissipation relation to
determine $\chi$. Instead, we generalize arguments for the RFIM by Schwartz and
Soffer \cite{schwartz1985exact} to express the magnetic susceptibility for the RFPM
in a different form. Consider the Hamiltonian of the RFPM in a uniform external
magnetic field $H^\alpha$,
\begin{equation}
  \label{hamilt_H}
  \mathcal{H}=-J\sum_{\left<ij\right>}\delta_{s_i,s_j}-
  \sum_i\sum_{\alpha=0}^{q-1}(h_{i}^{\alpha}+H^\alpha)\delta_{s_i,\alpha}.
\end{equation}
Then the ($q$-component) magnetic susceptibility is defined as
\begin{equation}
 \chi^\mu= \lim_{H^\mu\to 0} \frac{1}{N}\left[ \frac{\partial \langle M^\mu\rangle}{\partial H^\mu}\right]_\mathrm{av},
\end{equation}
where $\langle M^\mu\rangle$ denotes the thermal average of the total magnetization,
and $N$ is the number of spins. Defining
$ \bar{h} ^\alpha_i = h_i^\alpha + H^\alpha$, the disorder average
$[\cdots]_{\rm av}$ in above equation can be expressed as
\begin{equation}
  \left[ \frac{\partial \langle M^\mu\rangle}{\partial H^\mu}\right]_\mathrm{av} =
  \int\d\{\bar{h}_i^\alpha\} P(\{\bar{h}_i^\alpha\})  \frac{\partial \langle
    M^\mu\rangle_{\bar{h}^\alpha_i}}{\partial H^\mu}.
  \label{eq:schwartz1}
\end{equation}
Given that
\[
 \frac{\partial \langle M^\mu\rangle}{\partial H^\mu} = \sum_{i,\alpha} \frac{\partial \langle
 M^\mu\rangle}{\partial \bar{h}^\alpha_i} \frac{\partial \bar{h}^\alpha_i}{H^\mu} = \sum_i \frac{\partial \langle
 M^\mu\rangle}{\partial \bar{h}^\mu_i},
\]
partial integration applied to Eq.~\eqref{eq:schwartz1} yields
\begin{equation}
   \left[ \frac{\partial \langle M^\mu\rangle}{\partial H^\mu}\right]_\mathrm{av}= -  \sum_i \int\d\{\bar{h}_i^\alpha\}
  \frac{\partial P(\{\bar{h}_i^\alpha\})}{\partial  \bar{h}^\mu_i} \langle M^\mu\rangle_{\{ \bar{h}^\alpha_i\}}.
    \label{eq:schwartz}
\end{equation}
Since $P(\bar{h}_i^\alpha)$ is a normal distribution of mean $H^\alpha$ and variance
$\Delta$, we have
\[
  \frac{\partial P(\bar{h}_i^\mu)}{\partial \bar{h}_i^\mu} =
  -\frac{\bar{h}_i^\mu-H^\mu}{\Delta^2} P(\bar{h}_i^\mu).
\]
Using this in \eqref{eq:schwartz}, we hence find for the susceptibility,
\begin{equation}
\label{eq:chi}
 \chi^\mu = \lim_{H^\mu\to 0}   \frac{1}{N}\left[ \frac{\partial \langle M^\mu\rangle}{\partial H^\mu}\right]_\mathrm{av} =
  \frac{1}{\Delta^2}\left[\langle m^\mu\rangle \sum_i h_i^\mu\right]_\mathrm{av},
\end{equation}
where $m^\mu = M^\mu/N$ is the magnetization per spin in the state $\mu$.  
This form for $\chi^\mu$ provides the correct susceptibility as long as spontaneous
symmetry breaking occurs between the $q$ different states. While this is the case in
the thermodynamic limit, a suitably modified approach is necessary for finite
systems.  An explicit symmetry breaking in this case can be achieved by applying a
small external field that must be, however, sufficiently strong to actually break the
symmetry. From the Hamiltonian \eqref{hamilt_H}, if we look at the typical scale of
energy contributions due to a constant external field $H$ and the random fields of
strength $\Delta$, we see that an external field $H \gtrsim \Delta N^{-1/2}$ is
sufficient to break the symmetry (note $N=L^3$). In Table~\ref{parameter}, we present
the values of the constant external field $H$ for different lattice sizes $L$ chosen
to break the symmetry, such that the susceptibility $\chi$ exhibits a maximum. These
fields are applied to spin state 1 (i.e., $\mu =1$) and $\chi$ is computed from
Eq.~\eqref{eq:chi} for different $L$ and $\Delta$ after averaging over a large number
of disorder realizations $N_{\rm samp}$ as listed in Table~\ref{parameter}. Since
$\chi$, as defined in \eqref{eq:chi}, amounts to a correlation measure among the
magnetization and the total random field over the distribution of the latter, we find
that no reliable extrapolation in $n$ is possible for this quantity. We hence study
it for our largest finite $n = 100$.

\begin{figure}
\begin{center}
\includegraphics[width=\columnwidth]{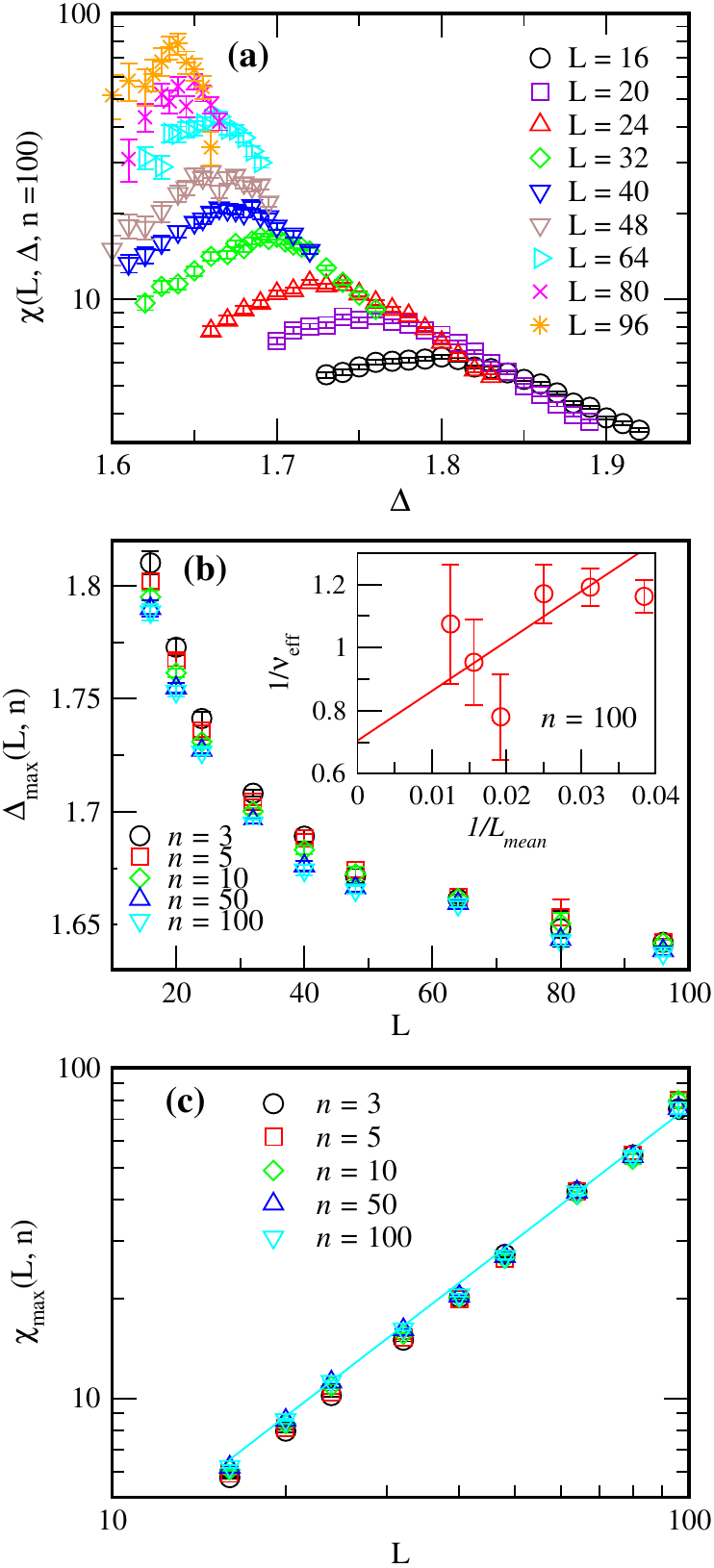}
\caption{(a) $\chi(L,\Delta)$ (on a semi log scale) for $n=100$ as a function of
  $\Delta$ and for all system sizes $L$ as specified. (b) $\Delta_{\rm \max}$, the
  location of the maxima of $\chi$, as a function of $L$ and for different
  $n$. Inset: effective exponent $1/\nu_{\rm eff}$ as a function of $1/L_{\rm mean}$
  for $n=100$. The solid line is a linear fit, yielding an extrapolated estimate
  $1/\nu=0.704(43)$ of the exponent, consistent with the corresponding estimate from
  the FSS of $[m]_{\rm av}(L,\Delta,n=100)$, see Table~\ref{mag_exp}.  (c) Maximum
  $\chi_{\max}(L)$ of the susceptibility against $L$ (on a log-log scale) for
  different $n$. The line corresponds to a power-law fit of the form
  $\chi_{\max}(L) \sim L^{\gamma/\nu}$ with an estimate of $\gamma/\nu=0.343(9)$ to
  the data for $\chi_{\max}(L)$ at $n=100$.  }
\label{chi}
\end{center}
\end{figure}

\begin{table}
\centering
\caption{Estimates of $\Delta_\mathrm{c}$, $1/\nu$, and $\gamma/\nu$ from fits of the
  locations $\Delta_{\rm \max}(L,n)$ and the heights $\chi_{\max}(L,n)$ of the
  susceptibility. $Q_1$ denotes the quality of the fit for the data of
  $\Delta_{\rm \max}(L,n)$, whereas $Q_2$ is the quality of the fit for the data of
  $\chi_{\max}(L,n)$.}
\begin{ruledtabular}
\begin{tabular}{ c  c  c  c c c  c}
 $n$&$\Delta_\mathrm{c}$&$1/\nu$&$\gamma/\nu$&$Q_1$&$Q_2$ \\ \hline
 3&1.625(6)&1.058(99)&1.401(10)&0.08&0.03 \\ 
 5&1.622(5)&1.065(85)&1.375(10)&  0.04&0.06 \\ 
 10&1.620(4)&1.064(77)&1.349(10)&0.04&0.02\\ 
 50&1.618(4)&1.054(89)&1.343(9)&0.05&0.05 \\ 
100&1.617(4)&0.963(67)&1.343(9)&0.05&0.03\\
 \end{tabular}
   \end{ruledtabular} 
\label{chi_exp}
\end{table}



Figure~\ref{chi}(a) shows $\chi(L,\Delta)$ as a function of $\Delta$ for $n=100$ and
all lattice sizes $16\le L \le 96$. A clear maximum in $\chi(L,\Delta)$ can be seen,
and the maxima move to lower $\Delta$ with growing $L$; further, the height of the
maximum continuously grows with $L$, signaling that $\chi$ has a divergent behavior.
To analyze this divergence, we fit a parabola near the maximum in order to obtain the
location $\Delta_{\max, \chi}$ and the height $\chi_{\max}$ of the susceptibility
maximum. Figures~\ref{chi}(b) and (c), respectively, show $\Delta_{\max, \chi} (L)$
and $\chi_{\max}(L)$ as a function of $L$ and for different $n$. Note that
$\chi_{\max} (L)$ is plotted on a log-log scale. From the FSS ansatz
\begin{equation}
\label{fss_chi}
 \chi(L,\Delta) =L^{\gamma/\nu} \widetilde{\mathcal{\chi}}\left[(\Delta-\Delta_\mathrm{c})L^{1/\nu}\right]
\end{equation}
we deduce that
\begin{equation}
\label{deltachi}
 \Delta_{\max,\chi}(L)\approx\Delta_\mathrm{c}+a_1 L^{-1/\nu},
\end{equation}
and 
\begin{equation}
 \chi_{\max}(L) \sim L^{\gamma/\nu}.
\end{equation}
Fits of the form \eqref{deltachi} to our data for $\Delta_{\max, \chi} (L)$ work
marginally well and lead to the parameters $\Delta_\mathrm{c}$ and $1/\nu$ that are
compiled in Table~\ref{chi_exp}. If we compare these values to those from
magnetization (Table~\ref{mag_exp}) and specific heat (Table~\ref{sp_heat_exp}), we
see that the values of $\Delta_\mathrm{c}$ are consistent but the values of $1/\nu$
are appreciably larger than the previous estimates, strongly suggesting the presence
of corrections to scaling. Unfortunately, including a correction-to-scaling term,
i.e., $\Delta_{\max, \chi} (L)=\Delta_\mathrm{c}+a_1 L^{-1/\nu}(1+a_2L^{-\omega})$
leads to very unstable fits and hence a reliable analysis is not possible. Instead,
we again reverted to the concept of effective exponents analogous to
Eq.~\eqref{eq:nu_eff}. A linear extrapolation of the estimates
$1/\nu_\mathrm{eff}(L)$ for $n=100$ as shown in the inset of Fig.~\ref{chi}(b) for
this case yields $1/\nu = 0.704(43)$, which is consistent with the corresponding
$n=100$ estimate for the order parameter scaling.

\begin{figure}[tb]
\begin{center}
\includegraphics[width=\columnwidth]{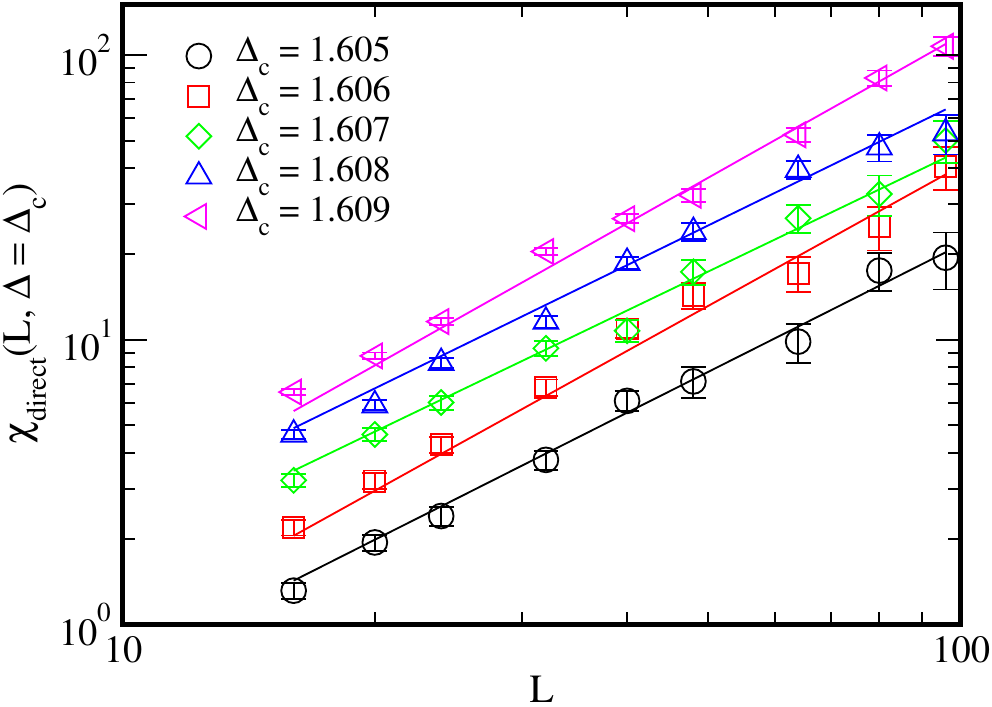}
\caption{FSS behavior of $\chi_ {\rm direct} $ (on a double-log scale), evaluated at
  several $\Delta_\mathrm{c}$ for the largest $n=100$. For a better view, the data
  for different $\Delta_\mathrm{c}$ are shifted relative to each other through
  multiplication by a constant factor. The lines are fits of the power-law
  $\chi_ {\rm direct} (L, \Delta_\mathrm{c})\sim L^{\gamma/\nu}$ for all
  $L\ge L_{\min}$ to the data. The resulting exponents $\gamma/\nu$ from different
  choices of $L_{\min}$ are collected in the Table ~\ref{gamma_exp}.}
 \label{chi_deltac}
\end{center}
\end{figure}

Turning to $\chi_{\max}(L)$, we fit the power-law
$\chi_{\max}(L) \sim L^{\gamma/\nu}$ to the data of Fig.~\ref{chi}(c) and collect the
results for the exponent $\gamma/\nu$ alongwith the (rather marginal) fit qualities
$Q_2$ in Table~\ref{chi_exp}. Here, we do not find any clean signature of a
correction to scaling; the fit results vary only slightly on excluding small $L$, and
within error bars they are consistent. Given the experience with the peak locations
we were nevertheless skeptical about the apparent absence of corrections, and we
hence conducted additional analyses of the susceptibility $\chi$ at fixed $\Delta$
and \emph{without} the explicit symmetry breaking through a small external field. As
our best estimate for the critical field strength is $\Delta_\mathrm{c}=1.607(2)$, we
conducted such simulations in the range $\Delta_\mathrm{c}=[1.605,1.609]$. The
resulting quantity, which we denote as $\chi_ {\rm direct}$, is shown at different
estimates for $\Delta_\mathrm{c}$ in Fig.~\ref{chi_deltac}. (The data have been
shifted vertically by multiplying a constant factor so that $\chi_ {\rm direct}$ at
different $\Delta_\mathrm{c}$ can be seen distinctly.) Power-law fits according to
Eq.~\eqref{fss_chi} work quite well. To identify any corrections to scaling, the
power law fits are performed on various ranges of $L\ge L_{\min}$.  The fit results
for the exponent $\gamma/\nu$ together with the fit qualities $Q$ are collected in
Table~\ref{gamma_exp}. Clearly, there is very little systematic variation in the
resulting estimate of $\gamma/\nu$ on variation of $L_{\min}$. We hence put forward
the fits for the complete range of $L$, i.e., for $L_{\min}=16$.  Such fits at
various $\Delta_\mathrm{c}$ are shown as solid lines in Fig.~\ref{chi_deltac}, and
the corresponding $\gamma/\nu$ estimates for different $\Delta_\mathrm{c}$ as found
in Table~\ref{gamma_exp} are consistent with each other. We quote
$\gamma/\nu=1.467(53)$ corresponding to the value at $\Delta_\mathrm{c}=1.607$ and
for $L_{\min}=16$ as our final estimate. Clearly, there is a noticeable difference in
the values of $\gamma/\nu$ estimated through this direct approach from the ones of
the $\chi$ estimates including the symmetry-breaking field, but we believe that the
unbroken estimates are the more reliable ones. This is in broad agreement with other
experiences concerning the connected susceptibility in critical phenomena, which is
often found to suffer from significant corrections, see, e.g.\
Refs.~\onlinecite{janke:04c,akritidis:23}.

\begin{table}[tb]
\centering
\caption{Estimates of the exponent $\gamma/\nu$ after fitting the form
  $\chi(L,\Delta_\mathrm{c}) \sim L^{\gamma/\nu}$ for all five $\Delta_\mathrm{c}$ to
  the corresponding data. The fits are performed to all $L\ge L_{\min}$ with the
  cut-offs $L_{\min}=16$, $20$, and $24$, respectively.}
\begin{ruledtabular}
\begin{tabular}{ c  c  c c c c c }
 
$\Delta_\mathrm{c}$ & \multicolumn{2}{c}{$L_{\min}=16$} & \multicolumn{2}{c}{$L_{\min}=20$} &\multicolumn{2}{c}{$L_{\min}=24$}  \\
\cline{2-3} \cline{4-5} \cline{6-7}  
&$\gamma/\nu$&$Q$&$\gamma/\nu$&$Q$&$\gamma/\nu$&$Q$\\ \hline
1.605& 1.542(52)& 0.62 &1.529(64)&0.52&1.557(80)&0.43\\ 
1.606&1.562(51)& 0.22 &1.560(62)&0.16&1.559(78)&0.11\\ 
1.607& 1.467(53)& 0.63 &1.447(66)& 0.54&1.463(84)& 0.42 \\ 
1.608& 1.488(50)&0.45&1.531(62)& 0.49&1.499(79)& 0.42\\
1.609& 1.556(48)&0.72&1.577(59)& 0.66&1.558(74)& 0.57\\
\end{tabular}
\end{ruledtabular}    
 \label{gamma_exp}
\end{table}

Finally, in Table~\ref{tab:exponents}, we present all our extrapolated results for
the critical exponents $\nu$, $\alpha$, $\beta$, $\gamma$, and $\bar \gamma$. The
exponents $\eta$, $\bar \eta$, and $\theta$ are obtained using the relations
$\gamma=\nu(2-\eta)$, $\bar \gamma=\nu(4-\bar\eta)$, and $\theta=2-\bar
\eta+\eta$. With these estimates, one can check the validity of the Rushbrooke
equality $\alpha+2\beta+\gamma=2$ and the {\it modified} hyperscaling relation
$2-\alpha=\nu (d-\theta)$.
\cite{schwartz1991breakdown,grinstein1976ferromagnetic,fisher1986scaling} It can be
easily inspected from Tables~\ref{tab:exponents} that both relations are well
satisfied (within error bars). The so-called two-exponent scaling scenario predicts
$\bar \eta = 2\eta$. \cite{schwartz1985random,schwartz1991missing} This is also
compatible with our results for $q=4$ as we find a very marginal value of the
difference $2\eta-\bar{\eta} = 0.013(53)$, indicating that there are only two
independent critical exponents. In Table~\ref{tab:exponents}, we also show the
exponents from our previous study for $q=3$ RFPM \cite{kumar2022critical} and the
recent estimates for the three-dimensional RFIM \cite{fytas2017restoration}.

\begin{table}[tb!]
    \centering
    \caption{Critical exponents of the $q=4$ RFPM as compared those of the $q=3$ RFPM
      \cite{kumar2022critical} and the RFIM (the $q=2$ case)
      \cite{fytas2017restoration}. All exponents are for dimension $d=3$ and the
      numbers in the parenthesis denote the error bars in the last significant
      digits.}
    \label{tab:exponents}
\begin{ruledtabular}
\begin{tabular}{crrr}
         &   RFIM & $q=3$ RFPM& $q=4$ RFPM\\ \hline
         $\nu$ & 1.38(10) &  1.383(8) &  1.55(2)\\
         $\alpha$ & -0.16(35) & -0.082(28)&  -0.31(12) \\
         $\beta$ & 0.019(4) & 0.0423(32)&  0.061(2) \\
         $\gamma$ & 2.05(15) & 2.089(84) & 2.274(87) \\
         $\bar\gamma$ & 4.10(6) & 4.07(3) & 4.53(6) \\
         $\eta$ & 0.5139(9) & 0.49(6) & 0.533(53) \\
         $\bar{\eta}$ & 1.028(2) & 1.060(3) & 1.080(3)\\
         $\theta$ & 1.487(1) & 1.43(6) & 1.45(6)\\
         $\alpha+2\beta+\gamma$ & 2.00(31) & 2.08(9)& 2.09(15) \\  
    \end{tabular}
 
\end{ruledtabular}
\end{table}

\section{Summary and Discussion}
\label{s4}

Based on a recently developed tool-chain of quasi-exact ground-state calculations for
Potts models with random fields, we have investigated the four-state random-field
Potts model on the cubic lattice with a focus on its zero-temperature critical
behavior. Our approach is based on the application of graph-cut methods, which are
known to be exact for two-state (Ising like) systems, to an embedding of two-state
variables into the $q$ states of the Potts system
\cite{bvz,kumar2018approximate}. Combining this method with a systematic
extrapolation technique based on $n$ repetitions of such approximate ground-state
calculations provides access to zero-temperature behavior of the system while
avoiding the exponentially growing effort for any exact approach applied to this NP
hard optimization problem \cite{kumar2023quasi}.

Akin to the situation for other random-field systems, the location of the disordered
fixed point at temperature $T=0$ implies that such ground-state calculations also
describe the critical behavior at non-zero temperatures. While for the pure Potts
system discontinuous transitions occur for all $q \ge 3$, the transition order for
the random-field problem is not known. In a previous study we showed that for $q=3$
states the RFPM retains a continuous transition \cite{kumar2022critical}. In the
present work we generalized these considerations to the four-states problem. We have
calculated the physical quantities, such as the magnetization, Binder cumulant, bond
energy, specific heat, as well as the connected and disconnected susceptibilities, on
various system sizes up to $96^3$. To average these quantities, the simulations were
performed for a large number of disorder samples and each disorder sample is further
simulated for $n=100$ initial conditions to facilitate the extrapolation to the
quasi-exact limit $n\to \infty$. Employing a comprehensive finite-size scaling
analysis, we find that the first order transition of the pure system is softened
through the random fields to become continuous. Since the studied case of $T=0$
corresponds to the strongest disorder along the transition line, and at the weakest
disorder close to the pure case the transition is of first order, one should expect a
tricritical point at some intermediate disorder strength which, however, could only
be discovered using finite-temperature methods such as Monte Carlo simulations. This
structure of the phase diagram is rather different from the ones proposed in
Refs.~\onlinecite{gx,shapir84,ebjpcm} that had anticipated first-order transitions at
zero temperature.

Studying the finite-size scaling of the magnetic and energetic quantities, we extract
the full set of critical exponents of the model. Our estimates of various exponents
are broadly presented in Tables~\ref{mag_exp} to \ref{gamma_exp}. The last
Table~\ref{tab:exponents} lists the extrapolated estimate of all exponents. As a
sanity check, we test these exponent estimates against the Rushbrooke scaling law and
find it satisfied. It is clear that the $q=4$ estimates are distinct from those of
the $q=3$ RFPM as well as the RFIM, indicating that the $q=4$ RFPM lies in a
different universality class. In particular, the correlation length exponent as well
as the magnetic exponents appear to be clearly distinct from the cases $q=2$ and
$q=3$. The hyperscaling-violation exponent $\theta$, on the other hand, is consistent
within error bars between all three models, so it is conceivable that it is
independent of $q$ in this class of models \footnote{This is in contrast to the RFIM
  for different lattice dimensions, where $\theta$ is found to increase up to
  $\theta=2$ in $d=5$, see Ref.~\onlinecite{fytas2017restoration}}. The most subtle
determination for random-field problems concerns the specific-heat exponent
$\alpha$. Carefully taking scaling corrections into account, we here find a clearly
negative value $\alpha = -0.31(12)$, whereas the three-state model was found to be
closer to marginal with $\alpha = -0.082(28)$.

While these results imply a significant step forward in the understanding of the RFPM
in three dimensions, a lot remains to be desired. Is there a multicritical value
$q_\mathrm{c}$ such that the transitions become first order even at $T=0$ for
$q > q_\mathrm{c}$? How does the finite-temperature phase-diagram look like? Does the
behavior of the RFPM depend on the particular form of coupling the random fields to
the spins and on the random-field distribution? These and related questions form
interesting avenues for further studies.
 
\acknowledgments

MK and MW acknowledge the support of the Royal Society through a SERB-Newton
International fellowship (NIF$\backslash$R1$\backslash$180386) and the Newton International Fellowships Alumni Grants (AL$\backslash$221025, AL$\backslash$231045, and AL$\backslash$24100026). MK acknowledges the support of DST-SERB via a Ramanujan Fellowship (RJF/2022/00149). The bulk of
numerical simulations were done on the parallel compute cluster {\it Zeus} of
Coventry University.

\appendix

\section{Direct extrapolation of the correlation-length exponent}
\label{sec:appendix}

\begin{figure}
  \begin{center}
    \includegraphics[width=\columnwidth]{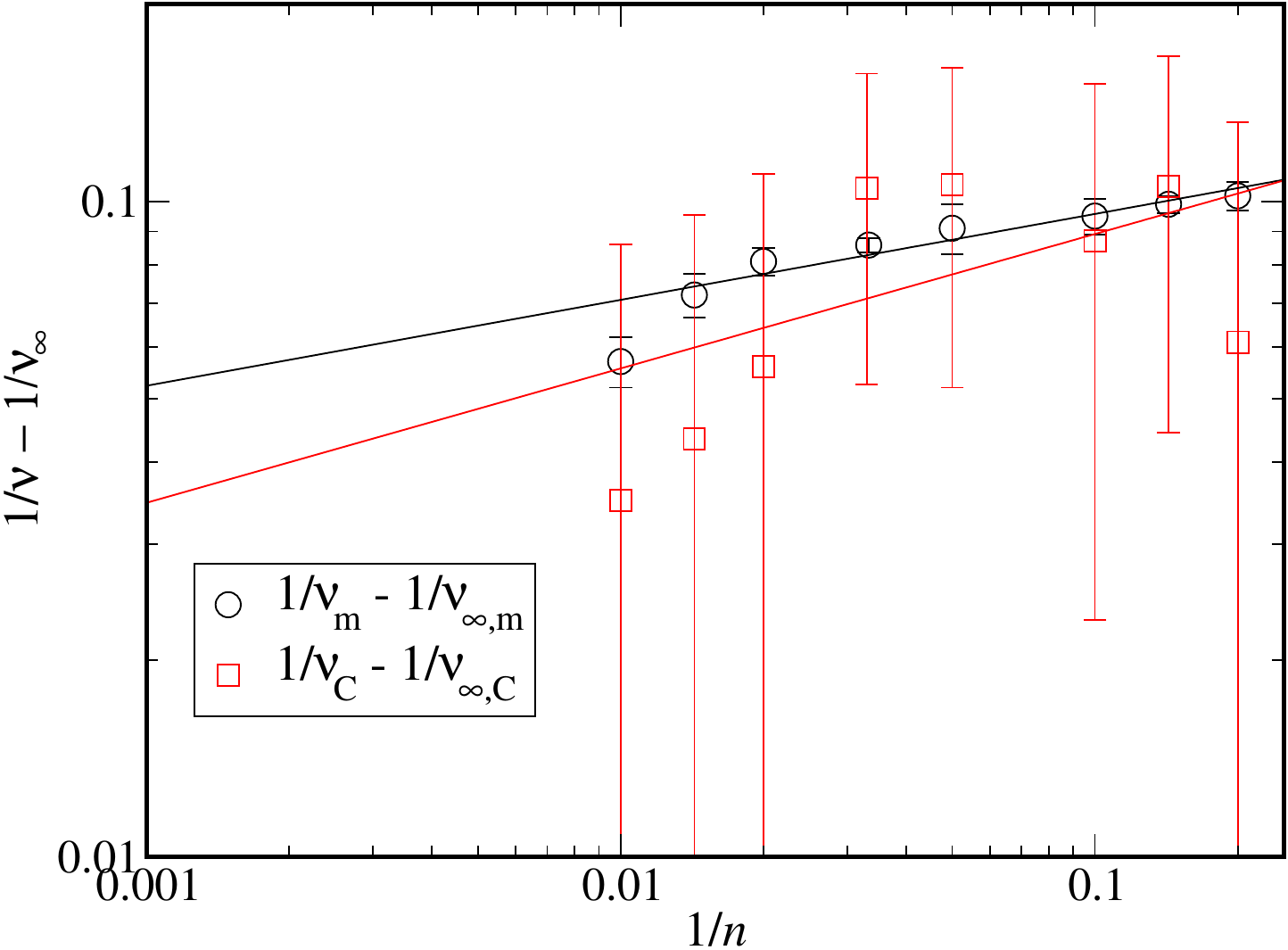}
    \caption{Extrapolation of the exponent estimates for $\nu^{-1}$ to the limit
      $n\to\infty$. The lines show the corresponding fits of the functional form
      \ref{nu_extrapolate_eq} to the data.}
    \label{nu_extrapolate_fig}
  \end{center}
\end{figure}

In the main text we compared the values of critical exponents determined from the
data for finite $n$ with those extracted from the data extrapolated for
$n\to\infty$. Here, we investigate the consistency of the two approaches for the
example of the disorder-averaged magnetization $[m]_{\rm av}(L,\Delta,n)$ and the
peak positions of specific-heat, \emph{viz.}\ $\Delta_{\rm \max, C}(L,n)$ [see
Tables~\ref{mag_exp} and \ref{sp_heat_exp}]. To this end, we considered the
possibility of a power-law form for the finite-$n$ deviations of the (inverse)
correlation length exponent $1/\nu$,
\begin{equation}
\label{nu_extrapolate_eq}
 \nu^{-1}(n)= \nu^{-1}_{\infty}+a n^{-c},  
\end{equation}
where $\nu^{-1}_\infty$ is the extrapolated value of $\nu^{-1}$ in the limit
$n\to\infty$. Attempting such fits for the estimates $\nu_m^{-1}(n)$ from the
magnetization data collapse and $\nu_C^{-1}$ for the scaling of the specific-heat
peak locations, we are not able to arrive at stable results if leaving all three
parameters to vary.  The fit routine produces negative extrapolated values for both
$1/\nu_{\infty,m}$ and $1/\nu_{\infty,C}$ with a small power-law exponent $c$, which
are nonphysical. If, on the other hand, we fix $\nu_{\infty,m}^{-1}=0.645$ and
$\nu_{\infty,C}^{-1}=0.686$ from Tables~\ref{mag_exp} and \ref{sp_heat_exp},
respectively, we find acceptable fits with $b = 0.1294(66)$, $c = 0.131(17)$ with
$\chi^2/{\rm d.o.f}=1.86$ from $1/\nu_m$, and $b = 0.143(55)$, $c = 0.206(26)$ with
$\chi^2/{\rm d.o.f}=0.23$ from $1/\nu_C$. As the residuals
$\nu^{-1}(n)-\nu_\infty^{-1}$ shown in Fig.~\ref{nu_extrapolate_fig} reveal, the
signal-to-noise ratio in these corrections is too small for a reliable non-linear
three-parameter fit. We hence conclude that the extrapolations of the exponent
estimates for finite $n$ are consistent with the $n\to\infty$ result, but an
extrapolation of the observable quantities to $n\to\infty$ is the more reliable
strategy as compared to an extrapolation of the exponent estimates themselves.

\bibliography{ref.bib}

\end{document}